\documentclass{article}
\usepackage[utf8]{inputenc}
\usepackage{xcolor}
\usepackage{graphicx}
\usepackage{subcaption}
\usepackage{float}
\usepackage[sectionbib,authoryear]{natbib}
\usepackage{setspace}
\usepackage{hyperref}
\usepackage{booktabs}
\usepackage{booktabs} 
\usepackage{amsmath}  
\usepackage{amssymb}  
\usepackage{makecell} 
\usepackage{array}
\usepackage{comment}
\newcommand{\keywords}[1]{%
  \par\noindent\textbf{Keywords:} #1\par
}

\usepackage{amsmath,amsfonts,amssymb,amsthm, bm,authblk,mathtools}
\usepackage{booktabs}
\usepackage{siunitx}
\usepackage{graphicx}
\usepackage{placeins}
\usepackage{adjustbox}
\usepackage{multirow}

\usepackage{longtable}
\title{Zero-inflated generalized linear latent variable models for microbiome compositional data}

\author[1]{Wenqi Tang\thanks{Corresponding author: wenqi.t.tang@jyu.fi}}
\author[1]{Pekka Korhonen}
\author[1]{Jenni Niku}
\author[2]{Klaus Nordhausen}
\author[1]{Sara Taskinen}

\affil[1]{Department of Mathematics and Statistics, University of Jyväskylä, Finland}
\affil[2]{Department of Mathematics, University of Helsinki, Finland}

\date{}

\begin{document}

\maketitle

\begin{abstract}
\textcolor{black}{
In this paper, we incorporate the Dirichlet--multinomial distribution into the generalized linear latent variable model (GLLVM) framework to accommodate high-dimensional, overdispersed compositional count data and capture complex covariation among taxa with a low-rank structure. Advances in high-throughput sequencing technologies have generated large amounts of microbiome count data, stimulating research into their associations with human diseases and traits. Such data are compositional and often highly sparse for various reasons, posing challenges for statistical analysis. Conventional zero-replacement approaches ignore the underlying causes of zeros, further complicating the analysis. We therefore propose a zero-inflated Dirichlet--multinomial latent variable model (ZIDM-LVM). Since the marginal likelihood has no closed-form expression,
we develop a variational approximation (VA) approach for
efficient parameter estimation.
Simulation studies and applications to two microbiome
datasets are presented to evaluate the performance
of the proposed method. }
\end{abstract}


\section{Introduction}

Recent advances in high-throughput sequencing (HTS) technologies have enabled large-scale characterization of microbial communities through taxon count data, facilitating the study of associations between microbial composition and biological and environmental factors \citep{Knight2018,Gloor2017}. Sequencing generates a finite number of reads for each sample that are distributed among taxa, resulting in compositional count data whose components are constrained by a fixed total and are therefore not independent \citep{Gloor2017}.

For compositional data, \citet{Aitchison1982} introduced the Aitchison geometry on the constrained simplex, rather than the conventional Euclidean geometry. This work also established the foundations of compositional data analysis (CoDA) through a log-ratio framework that considers relative rather than absolute abundances and analyzes ratios between components. Within this framework, log-ratio transformations, such as the centered log-ratio (CLR) transformation, map compositions from the constrained simplex to Euclidean space, allowing standard multivariate methods to be applied. However, many CoDA methods require strictly positive components and therefore cannot be applied directly when zeros are present.

Microbiome count data are often highly sparse, with zeros occurring frequently. These zeros may arise from different sources. A taxon undetected due to limited sequencing depth gives a sampling zero, whereas the true absence of a taxon gives a structural zero. These different sources of zeros imply that the two types carry different information and require different approaches to handling them. Sampling zeros do not necessarily represent true absence and can be replaced with small positive values before analysis, while structural zeros carry information about true absence and are often addressed by statistical models incorporating a structural-zero component. \citet{Tang2026} summarized commonly used approaches for handling
sampling and structural zeros in high-dimensional compositional
count data.

The multinomial (MN) distribution provides a natural starting point for modeling compositional count data by conditioning on the total number of reads observed in each sample. However, the MN distribution cannot accommodate overdispersion arising from greater between-sample variation in microbial count data. Several extensions have been proposed to address overdispersion. The multiplicative multinomial distribution \citep{altham2012:mmn} allows pairwise interactions between taxa but involves $p(p-1)/2$ dispersion parameters and a computationally demanding normalizing constant. The Conway--Maxwell multinomial distribution \citep{morris2020:commn} similarly involves a difficult normalizing term but requires only one dispersion parameter per observational unit. The random-clumped multinomial (RCM) distribution \citep{neerchal93:rcm} introduces overdispersion through a clumping mechanism, where some trials within a multinomial draw follow the outcome of another trial. The classical RCM allows one clump per draw, while extensions allow multiple clumps \citep{banerjee1999:mrcm}.

Another distribution that naturally accommodates compositional count data is the Dirichlet--multinomial (DM) distribution. It extends the MN model by allowing the underlying composition to vary between samples, thereby accommodating overdispersion.

Extensions of both the MN and DM distributions have incorporated zero inflation (ZI) to account for the high prevalence of zeros in compositional count data. A ZI generalized DM model based on a ZI Beta stick-breaking representation was proposed in \citet{tang:2018:zigdm}. For high-dimensional HTS data, a ZI logistic-normal multinomial model with a low-rank latent representation was developed in \citet{Zeng2023}. A Bayesian ZI DM regression model based on a Gamma representation of the Dirichlet distribution was introduced in \citet{Koslovsky2023}. More recently, zero- and $N$-inflated latent variable models based on the MN and DM distributions using Gamma augmentation were proposed in \citet{menezes2025:zanim}. These developments suggest that combining the DM distribution with ZI provides a flexible approach to accommodating both excess zeros and overdispersion beyond that associated with excess zeros.

In this work, we incorporate the MN and DM distributions into the generalized linear latent variable model (GLLVM) framework to model compositional count data. GLLVMs accommodate high-dimensional multivariate responses by combining taxon-specific covariate effects with a low-dimensional latent structure that captures residual dependence among taxa \citep{Skrondal2004,Warton2015,Hui2015,Niku2017}. To further accommodate excess zeros, we incorporate a ZI component representing structural zeros, leading to two models: the zero-inflated multinomial latent variable model (ZIMN-LVM) and the zero-inflated Dirichlet--multinomial latent variable model (ZIDM-LVM). For efficient likelihood-based estimation of these high-dimensional latent variable models, we use marginal maximum likelihood with variational approximation \citep{Hui2017,Niku2019,Zeng2023,Korhonen2023}.

The remainder of the paper is organized as follows.
Section~\ref{sec:model_framework} presents the GLLVM framework and
variational approximation. Section~\ref{sec:zidm} introduces the proposed ZIDM-LVM and
develops the corresponding variational approximation. Section~\ref{sec:simulation} describes the simulation study
and evaluates the performance of the proposed methods.
Section~\ref{sec:real_data_analysis} presents applications to microbiome
data, and Section~\ref{sec:discussion} concludes the paper with a discussion.
 \textcolor{black}{The formulation and variational approximation of the ZIMN-LVM
are provided in Appendix~\ref{app:zimn}. Appendix~\ref{app:zimn_dgm}
describes the ZIMN-LVM data-generating mechanism and reports the
corresponding simulation results, Appendix~\ref{app:additional_simulation}
compares conventional and classification VA, and
Appendix~\ref{app:notation} summarizes the notation and abbreviations used throughout the paper.}

\section{Model framework}
\label{sec:model_framework}
\subsection{Generalized linear latent variable models}

Generalized linear latent variable models (GLLVMs) provide a flexible
framework for jointly modeling multivariate responses by extending
generalized linear models with a low-dimensional latent variable
structure \citep{Skrondal2004,Warton2015,Hui2015}. An important
advantage of GLLVMs is their ability to accommodate response data with
different distributional characteristics through an appropriate choice
of conditional distribution and link function, including
presence--absence data, counts, biomass, cover, and other continuous or
semi-continuous responses
\citep{Warton2015,Korhonen2023,Korhonen2025}. This flexibility also
allows the GLLVM framework to be extended to compositional count data
by specifying conditional response distributions that explicitly
account for the compositional sampling structure.

Let $y_{ij}$ denote the response of taxon $j=1,\ldots,p$ in sample
$i=1,\ldots,n$. Conditional on a $d$-dimensional latent variable
$\mathbf u_i$, the responses are assumed to be independent, with
$y_{ij}\mid\mathbf u_i\sim F(\mu_{ij},\boldsymbol{\vartheta})$ and
$\mathbf u_i\sim N_d(\mathbf 0,\mathbf I_d)$, where
$\mu_{ij}=E(y_{ij}\mid\mathbf u_i)$ and $\boldsymbol{\vartheta}$ denotes
possible additional response-specific parameters, such as dispersion
or zero-inflation parameters for count-data models
\citep{Hui2015,Korhonen2025}. The conditional mean is related to the
linear predictor through
\begin{equation*}
\label{eq:gllvm_general}
g(\mu_{ij})
=
\eta_{ij}
=
\alpha_i
+
\beta_{0j}
+
\mathbf x_i^\top\boldsymbol{\beta}_j
+
\mathbf u_i^\top\boldsymbol{\gamma}_j,
\end{equation*}
where $g(\cdot)$ is a known link function,
$\beta_{0j}$ is a taxon-specific intercept,
$\mathbf x_i$ is a vector of observed covariates, and
$\boldsymbol{\beta}_j$ contains the corresponding taxon-specific
regression coefficients. The sample-specific effect $\alpha_i$ may be
treated as either fixed or random and can be used to account for
differences in sampling intensity or total abundance among samples
\citep{Warton2015,Niku2019,Korhonen2023}.

The vector $\boldsymbol{\gamma}_j$ contains the taxon-specific latent
variable loadings. The latent variables $\mathbf u_i$ represent
unobserved factors underlying variation among samples, and the term
$\mathbf u_i^\top\boldsymbol{\gamma}_j$ captures residual dependence
among taxa that is not explained by the observed covariates
\citep{Warton2015,Niku2017}. Typically, the latent dimension $d$ is
chosen to be much smaller than the number of taxa $p$, providing a
parsimonious low-rank representation of the dependence structure among
taxa. For model-based ordination, the latent variables can be
interpreted as low-dimensional sample scores, with
$\boldsymbol{\gamma}_j$ describing the corresponding taxon loadings
\citep{Hui2015,Warton2015}.

GLLVMs can be estimated by maximum likelihood. Let
$\boldsymbol{\Psi}$ collect the model parameters and
$\mathbf u=(\mathbf u_1^\top,\ldots,\mathbf u_n^\top)^\top$ collect
the latent variables into vectors. Since the latent variables are unobserved, the
marginal log-likelihood is obtained by integrating them out,
\begin{equation}
\label{eq:gllvm_marginal_lik}
\ell(\boldsymbol{\Psi})
=
\log L(\boldsymbol{\Psi})
=
\log
\int_{\mathbb R^{nd}}
\left[
\prod_{i=1}^{n}
\left\{
\prod_{j=1}^{p}
f(y_{ij}\mid\mathbf u_i,\boldsymbol{\Psi})
\right\}
f(\mathbf u_i)
\right]
d\mathbf u .
\end{equation}

\subsection{Variational approximation}

The marginal log-likelihood in
\eqref{eq:gllvm_marginal_lik} generally does not have a closed-form
expression for non-normal responses and therefore requires an
approximation. We use variational approximation (VA), which replaces
the marginal log-likelihood by a variational lower bound
\citep{Hui2017,Niku2019,Korhonen2023}.

For GLLVMs, the variational lower bound can be obtained by introducing
a variational density $q(\mathbf u)$ for the latent variables
\citep{Hui2017}. The marginal log-likelihood can be written as
\begin{equation}
\label{eq:general_elbo}
\begin{aligned}
\ell(\boldsymbol{\Psi})
&=
\log
\int
f(\mathbf y,\mathbf u\mid\boldsymbol{\Psi})
\,d\mathbf u \\
&=
\int
q(\mathbf u)
\log
\left\{
\frac{
f(\mathbf y,\mathbf u\mid\boldsymbol{\Psi})
}{
q(\mathbf u)
}
\right\}
d\mathbf u +
\int
q(\mathbf u)
\log
\left\{
\frac{
q(\mathbf u)
}{
f(\mathbf u\mid\mathbf y,\boldsymbol{\Psi})
}
\right\}
d\mathbf u \\
&=
\operatorname{ELBO}(\boldsymbol{\Psi}\mid q)
+
\operatorname{KL}
\left\{
q(\mathbf u)
\,\|\,
f(\mathbf u\mid\mathbf y,\boldsymbol{\Psi})
\right\} \\
&\geq
\operatorname{ELBO}(\boldsymbol{\Psi}\mid q),
\end{aligned}
\end{equation}
where the second integral in \eqref{eq:general_elbo} is the
Kullback--Leibler (KL) divergence between $q(\mathbf u)$ and
$f(\mathbf u\mid\mathbf y,\boldsymbol{\Psi})$. Since the KL
divergence is non-negative, $\operatorname{ELBO}(\boldsymbol{\Psi}\mid q)$
provides a lower bound to the marginal log-likelihood and is referred
to as the evidence lower bound (ELBO). It follows that VA is concerned
with choosing a variational distribution that maximizes the lower
bound, or equivalently minimizes the corresponding KL divergence
\citep{Hui2017}.

For computational convenience, $q(\mathbf u)$ is restricted to a
parametric family
$q(\mathbf u;\boldsymbol{\xi})$,
where $\boldsymbol{\xi}$ denotes the variational parameters
\citep{Hui2017}. A
mean-field assumption is further used in the proposed models, under
which the variational distributions associated with distinct groups
of latent and augmented quantities are taken to be mutually
independent. The exact factorization and the corresponding
variational distributions depend on the model and are specified in
the following sections. Under the chosen parametric variational family, the ELBO can be
written as
$\operatorname{ELBO}(\boldsymbol{\Psi},\boldsymbol{\xi}\mid q)
=
\mathbb E_q[\log f(\mathbf y,\mathbf u\mid\boldsymbol{\Psi})]
-
\mathbb E_q[\log q(\mathbf u;\boldsymbol{\xi})]$.
The model and variational parameters are then estimated jointly by
\begin{equation}
\label{eq:general_va_estimation}
\begin{aligned}
(\widehat{\boldsymbol{\Psi}},
 \widehat{\boldsymbol{\xi}})
&=
\arg\max_{\boldsymbol{\Psi},\boldsymbol{\xi}}
\operatorname{ELBO}(\boldsymbol{\Psi},\boldsymbol{\xi}\mid q) \\
&=
\arg\max_{\boldsymbol{\Psi},\boldsymbol{\xi}}
\left\{
\mathbb E_q
\left[
\log f(\mathbf y,\mathbf u\mid\boldsymbol{\Psi})
\right]
-
\mathbb E_q
\left[
\log q(\mathbf u;\boldsymbol{\xi})
\right]
\right\}.
\end{aligned}
\end{equation}
Hence, $\widehat{\boldsymbol{\Psi}}$ provides the VA solution for
maximum likelihood estimation, while
$q(\mathbf u;\widehat{\boldsymbol{\xi}})$ approximates the
conditional distribution of the latent variables
\citep{Hui2017,Niku2019}.

\section{Zero-inflated Dirichlet--multinomial latent variable model}
\label{sec:zidm}
\subsection{Model formulation and augmented likelihood}
We first consider the multinomial model without zero inflation indicators. Let
$\mathbf Y_i=(y_{i1},\ldots,y_{ip})^\top$ denote the vector of taxon
counts for sample $i$, with $\sum_{j=1}^p y_{ij}=N_i$. We assume
\begin{equation*}
\mathbf Y_i\mid N_i,\boldsymbol{\rho}_i
\sim
\operatorname{Multinomial}(N_i,\boldsymbol{\rho}_i),
\end{equation*}
where $\boldsymbol{\rho}_i=(\rho_{i1},\ldots,\rho_{ip})^\top$
denotes the underlying composition, where
$\sum_{j=1}^p\rho_{ij}=1$. To accommodate overdispersion relative to the multinomial model, we
allow the underlying composition $\boldsymbol{\rho}_i$ to vary
according to a Dirichlet distribution,
\begin{equation*}
\label{eq:zidm_dirichlet}
\boldsymbol{\rho}_i
\sim
\operatorname{Dirichlet}
(\theta_{i1},\ldots,\theta_{ip}).
\end{equation*}

To incorporate the Dirichlet composition into the GLLVM framework, we
use its Gamma representation and model the Dirichlet concentration
parameters through the GLLVM linear predictor. Specifically, the
Dirichlet composition can be represented by normalizing independent
Gamma random variables,
\begin{equation*}
\label{eq:zidm_gllvm}
\begin{aligned}
\lambda_{ij}\mid\mathbf u_i
&\sim
\operatorname{Gamma}(\theta_{ij},1),
\qquad
\rho_{ij}
=
\frac{\lambda_{ij}}
{\sum_{k=1}^p\lambda_{ik}},
\\
\theta_{ij}
&=
\exp(\eta_{ij}),
\qquad
\eta_{ij}
=
\alpha_i+\beta_{0j}
+\mathbf x_i^\top\boldsymbol{\beta}_j
+\mathbf u_i^\top\boldsymbol{\gamma}_j,
\\
\mathbf u_i
&\sim
N_d(\mathbf 0,\mathbf I_d),
\qquad j=1,\ldots,p.
\end{aligned}
\end{equation*}
Here, $\theta_{ij}>0$ denotes the taxon-specific Dirichlet
concentration parameter, while $\lambda_{ij}>0$ denotes the
corresponding Gamma latent variable. Conditional on
$\boldsymbol{\rho}_i$, the counts follow the multinomial sampling
model,
\begin{equation}
\label{eq:zidm_multinomial}
\begin{aligned}
\mathbf Y_i\mid N_i,\boldsymbol{\rho}_i
&\sim
\operatorname{Multinomial}(N_i,\boldsymbol{\rho}_i),
\\
p(\mathbf Y_i;N_i,\boldsymbol{\rho}_i)
&=
\binom{N_i}{y_{i1},\ldots,y_{ip}}
\prod_{j=1}^p \rho_{ij}^{y_{ij}}.
\end{aligned}
\end{equation}
Together with the Dirichlet model for $\boldsymbol{\rho}_i$, this
defines the Dirichlet--multinomial GLLVM.
\textcolor{black}{Unlike conventional GLLVMs,
the taxon counts are not conditionally independent given the latent
variables $\mathbf u_i$, but follow a joint Dirichlet--multinomial
distribution.} Using the multinomial likelihood in \eqref{eq:zidm_multinomial}
together with the independent Gamma densities of $\lambda_{ij}$ gives
\begin{equation}
\label{eq:zidm_joint_lambda}
p(\mathbf y_i,\boldsymbol{\lambda}_i\mid\mathbf u_i)
=
\binom{N_i}{y_{i1},\ldots,y_{ip}}
\left(
\sum_{j=1}^p\lambda_{ij}
\right)^{-N_i}
\prod_{j=1}^p
\frac{
\lambda_{ij}^{y_{ij}+\theta_{ij}-1}
\exp(-\lambda_{ij})
}{
\Gamma(\theta_{ij})
}.
\end{equation}

The joint density in \eqref{eq:zidm_joint_lambda} presents two
computational difficulties. First, the normalization term
$\left(\sum_{j=1}^p\lambda_{ij}\right)^{-N_i}$ couples the
taxon-specific Gamma variables within each sample. Following
\citet{menezes2025:zanim}, we address this using a Gamma augmentation,
introducing
$\phi_i\mid\boldsymbol{\lambda}_i
\sim
\operatorname{Gamma}
\left(N_i,\sum_{j=1}^p\lambda_{ij}\right)$,
where the second parameter denotes the rate. Using the Gamma integral
identity,
\begin{equation*}
\left(
\sum_{j=1}^p\lambda_{ij}
\right)^{-N_i}
=
\frac{1}{\Gamma(N_i)}
\int_0^\infty
\phi_i^{N_i-1}
\exp\left\{
-\phi_i\sum_{j=1}^p\lambda_{ij}
\right\}
\,d\phi_i.
\end{equation*}
Second, the Gamma densities contain the reciprocal Gamma term
$\Gamma(\theta_{ij})^{-1}$, where
$\theta_{ij}=\exp(\eta_{ij})$ depends nonlinearly on the latent
variables. To facilitate variational approximation of this term, we
introduce an additional Pólya inverse-gamma (P-IG) auxiliary variable
$\omega_{ij}\sim\operatorname{P\text{-}IG}(0)$ following
\citet{he2022polyaig}. If $p_0(\omega)$ denotes the density of a
$\operatorname{P\text{-}IG}(0)$ random variable, then
\begin{align*}
\frac{1}{\Gamma(\theta_{ij})}
&=
\theta_{ij}\exp(\gamma_{\mathrm E}\theta_{ij})
\mathbb E_{\omega_{ij}}
\left[
\exp(-\theta_{ij}^{2}\omega_{ij})
\right] \\
&=
\theta_{ij}\exp(\gamma_{\mathrm E}\theta_{ij})
\int_0^\infty
\exp(-\theta_{ij}^{2}\omega)
p_0(\omega)\,d\omega,
\end{align*}
where $\gamma_{\mathrm E}=-\psi(1) \approx 0.57721$ is the Euler--Mascheroni constant.

Therefore, for the sampling component, combining the Gamma and P-IG
augmentations gives
\begin{align*}
p(y_{ij},\lambda_{ij}&,\phi_i,\omega_{ij}\mid\mathbf u_i)\\
&\propto
\theta_{ij}\exp(\gamma_{\mathrm E}\theta_{ij})
\lambda_{ij}^{y_{ij}+\theta_{ij}-1}
\exp
\left\{
-(1+\phi_i)\lambda_{ij}
-\theta_{ij}^{2}\omega_{ij}
\right\}
p_0(\omega_{ij}).
\end{align*}

We now introduce zero inflation for observed zeros through a latent
indicator $z_{ij}$, where $z_{ij}=1$ denotes the structural-zero
component and $z_{ij}=0$ denotes the multinomial sampling component,
\begin{equation*}
\label{eq:zidm_indicator}
z_{ij}
\sim
\operatorname{Bernoulli}(\zeta_j),
\qquad
P(z_{ij}=1)=\zeta_j,
\end{equation*}
with $0<\zeta_j<1$. When $y_{ij}=0$, the observation may therefore
arise either from the structural-zero component ($z_{ij}=1$) or as a
sampling zero from the multinomial component ($z_{ij}=0$).

\textcolor{black}{When $z_{ij}=0$, taxon $j$ belongs to the
Dirichlet--multinomial sampling component, with}
{\color{black}
\begin{equation*}
\label{eq:zidm_zi_lambda}
\lambda_{ij}\mid z_{ij},\mathbf u_i
\sim
\begin{cases}
\operatorname{Gamma}(\theta_{ij},1),
& z_{ij}=0,\\[3pt]
\delta_0,
& z_{ij}=1,
\end{cases}
\end{equation*}
}
\textcolor{black}{where $\theta_{ij}=\exp(\eta_{ij})$ and $\delta_0$
denotes a point mass at zero. Conditional on
$\mathbf z_i=(z_{i1},\ldots,z_{ip})^\top$, the resulting compositional
probabilities are therefore}
{\color{black}
\begin{equation}
\label{eq:zidm_zi_composition}
\rho^{0}_{ij}(\mathbf z_i)
=
\frac{
(1-z_{ij})\lambda_{ij}
}{
\sum_{k=1}^p(1-z_{ik})\lambda_{ik}
}.
\end{equation}
}
It can be seen that when a component is a structural zero,
$z_{ij}=1$ removes the corresponding taxon-specific intensity and
forces $\rho_{ij}(\mathbf z_i)=0$, whereas the remaining positive
intensities are renormalized to sum to one.

The two components can be represented jointly as
\begin{equation*}
\label{eq:zidm_joint_factor}
\left[
(1-\zeta_j)
\theta_{ij}\exp(\gamma_{\mathrm E}\theta_{ij})
\lambda_{ij}^{y_{ij}+\theta_{ij}-1}
\exp\left\{
-(1+\phi_i)\lambda_{ij}
-\theta_{ij}^{2}\omega_{ij}
\right\}
p_0(\omega_{ij})
\right]^{1-z_{ij}}
\left[
\zeta_j\delta_0(y_{ij})
\right]^{z_{ij}}.
\end{equation*}
\textcolor{black}{where the first component corresponds to the
Dirichlet--multinomial sampling component under the Gamma and P-IG
augmentations, while the second component represents a structural
zero, assigning probability one to $y_{ij}=0$ and
$\lambda_{ij}=0$.}

Accordingly, the complete-data
log-likelihood of the fully augmented ZIDM-LVM can be written as
\begin{align}
\label{eq:zidm_complete_loglik}
&\log \mathcal L_{\mathrm{ZIDM}}
\left(
\boldsymbol{\Psi};
\mathbf y,\boldsymbol{\lambda},\mathbf z,
\mathbf u,\boldsymbol{\phi},\boldsymbol{\omega}
\right)
\nonumber\\
={}&
\sum_{i=1}^n
\Bigg\{
(N_i-1)\log\phi_i
-
\log\Gamma(N_i)
+
\log(N_i!)
-
\sum_{j=1}^p\log(y_{ij}!)
+
\log p(\mathbf u_i)
\Bigg\}
\nonumber\\
&+
\sum_{i=1}^n
\sum_{j=1}^p
(1-z_{ij})
\Bigg[
\log(1-\zeta_j)
+
\eta_{ij}
+
\gamma_{\mathrm E}\theta_{ij}
\nonumber\\
&\hspace{25mm}
+
(y_{ij}+\theta_{ij}-1)\log\lambda_{ij}
-
(1+\phi_i)\lambda_{ij}
-
\theta_{ij}^{2}\omega_{ij}
+
\log p_0(\omega_{ij})
\Bigg]
\nonumber\\
&+
\sum_{i=1}^n
\sum_{j:y_{ij}=0}
z_{ij}\log\zeta_j .
\end{align}

\subsection{Variational approximation}

The latent and
augmented quantities in the ZIDM-LVM are the latent variables $\mathbf u$,
the Gamma augmentation variables $\boldsymbol{\phi}$, the
zero-inflation indicators $\mathbf z$, the Gamma latent variables
$\boldsymbol{\lambda}$, and the P-IG augmentation variables
$\boldsymbol{\omega}$. The variational lower bound for the ZIDM-LVM is
\begin{equation*}
\operatorname{ELBO}
(\boldsymbol{\Psi},\boldsymbol{\xi}\mid q)
=
\mathbb E_q
\left[
\log
p(
\mathbf y,
\boldsymbol{\lambda},
\mathbf z,
\mathbf u,
\boldsymbol{\phi},
\boldsymbol{\omega}
\mid\boldsymbol{\Psi})
\right]
-
\mathbb E_q
\left[
\log
q(
\mathbf u,
\boldsymbol{\phi},
\mathbf z,
\boldsymbol{\lambda},
\boldsymbol{\omega};
\boldsymbol{\xi})
\right].
\end{equation*}

Building on the mean-field variational framework introduced
above, we adopt a structured mean-field approximation that
retains conditional dependencies between the zero-inflation
indicators and the associated latent and augmented variables.
Specifically, we take
\begin{equation*}
q(
\mathbf u,\boldsymbol{\phi},\mathbf z,
\boldsymbol{\lambda},\boldsymbol{\omega})
=
\prod_{i=1}^n
q(\mathbf u_i)q(\phi_i)
\prod_{j=1}^p
q(z_{ij})
q(\lambda_{ij}\mid z_{ij})
q(\omega_{ij}\mid z_{ij}).
\end{equation*}
where
\begin{equation*}
\begin{gathered}
q(\mathbf u_i) \sim N_d(\mathbf a_i,\mathbf A_i),
\qquad
q(\phi_i) \sim \operatorname{Gamma}(N_i,S_i),\\[4pt]
q(z_{ij}) \sim
\operatorname{Bernoulli}(\widetilde{\pi}_{ij}),
\qquad y_{ij}=0.
\end{gathered}
\end{equation*}
Under $q(\phi_i)=\operatorname{Gamma}(N_i,S_i)$,
$\mathbb E_q(\phi_i)=N_i/S_i$. We define
$\boldsymbol{\xi}
=
\{\mathbf a_i,\mathbf A_i,S_i,\widetilde{\pi}_{ij}\}$
as the collection of variational parameters, and
$\boldsymbol{\Psi}
=
\{\alpha_i,\beta_{0j},\boldsymbol{\beta}_j,
\boldsymbol{\gamma}_j,\zeta_j\}$
as the model parameters.

For the Dirichlet concentration parameter
$\theta_{ij}=\exp(\eta_{ij})$, we include the second-order correction
induced by the variational covariance $\mathbf A_i$. Defining
$\widetilde{\eta}_{ij}
=
\alpha_i+\beta_{0j}
+\mathbf x_i^\top\boldsymbol{\beta}_j
+\mathbf a_i^\top\boldsymbol{\gamma}_j$
as the linear predictor evaluated at the variational mean
$\mathbf a_i$, we have
\begin{equation}
\label{eq:zidm_theta_moments}
\begin{aligned}
\widetilde{\theta}_{ij}
:=
\mathbb E_q(\theta_{ij})
&=
\exp
\left\{
\widetilde{\eta}_{ij}
+
\frac{1}{2}
\boldsymbol{\gamma}_j^\top
\mathbf A_i
\boldsymbol{\gamma}_j
\right\},
\\
\widetilde c_{ij}
:=
\left\{
\mathbb E_q(\theta_{ij}^{2})
\right\}^{1/2}
&=
\exp
\left\{
\widetilde{\eta}_{ij}
+
\boldsymbol{\gamma}_j^\top
\mathbf A_i
\boldsymbol{\gamma}_j
\right\}.
\end{aligned}
\end{equation}
For the P-IG augmentation variable $\omega_{ij}$, using
$\widetilde c_{ij}$ in \eqref{eq:zidm_theta_moments}, we take
\[
q(\omega_{ij}\mid z_{ij}=0)
=
\operatorname{P\text{-}IG}(\widetilde c_{ij}).
\]
Specifically, $\operatorname{P\text{-}IG}(c)$ is the exponential tilt
of $\operatorname{P\text{-}IG}(0)$,
$p_c(\omega)=Z_c\exp(-c^2\omega)p_0(\omega)$, with
$\log Z_c=\gamma_{\mathrm E} c+\log\Gamma(1+c)$.

For the zero-inflation indicator $z_{ij}$, we follow the partially
collapsed strategy of \citet{menezes2025:zanim} and first marginalize
the Gamma latent variable $\lambda_{ij}$ from the sampling component.
For $y_{ij}=0$, the resulting sampling contribution is proportional
to $(1-\zeta_j)(1+\phi_i)^{-\theta_{ij}}$, whereas the
structural-zero contribution is proportional to $\zeta_j$.
Taking expectations with respect to
$q(\mathbf u_i)q(\phi_i)$ gives
\begin{equation}
\label{eq:zidm_pi}
\widetilde{\pi}_{ij}
=
\begin{cases}
0,
& y_{ij}>0,\\[6pt]
\operatorname{logit}^{-1}
\left[
\log
\left(
\frac{\zeta_j}{1-\zeta_j}
\right)
+
\widetilde{\theta}_{ij}
\mathbb E_q\{\log(1+\phi_i)\}
\right],
& y_{ij}=0.
\end{cases}
\end{equation}

The expectation
$\mathbb E_q\{\log(1+\phi_i)\}$ in
\eqref{eq:zidm_pi} has no convenient closed form under
$q(\phi_i)=\operatorname{Gamma}(N_i,S_i)$.
Using a second-order Taylor expansion around
$\mathbb E_q(\phi_i)=N_i/S_i$, we use
\begin{equation*}
\label{eq:zidm_logphi_approx}
\mathbb E_q\{\log(1+\phi_i)\}
\approx
\log
\left(
1+\frac{N_i}{S_i}
\right)
-
\frac{
N_i/S_i^2
}{
2\left(1+N_i/S_i\right)^2
}.
\end{equation*}

\textcolor{black}{
It is worth noting that substituting the second-order
Taylor approximation into the original ELBO does not
necessarily preserve its lower-bound property.
However, since the Gamma variational distribution
$q(\phi_i)=\operatorname{Gamma}(N_i,S_i)$ becomes
increasingly concentrated as $N_i$ increases,
the approximation is expected to be accurate
for sufficiently large $N_i$.
}
For the Gamma latent variable $\lambda_{ij}$, conditional on membership
in the sampling component, the Gamma kernel gives
\begin{equation*}
\label{eq:zidm_qlambda}
q(\lambda_{ij}\mid z_{ij}=0)
=
\operatorname{Gamma}
\left(
y_{ij}+\widetilde{\theta}_{ij},
1+\frac{N_i}{S_i}
\right),
\end{equation*}
whereas $\lambda_{ij}=0$ when $z_{ij}=1$.

Combining the variational distributions and the corresponding
moments with the augmented complete-data likelihood in
\eqref{eq:zidm_complete_loglik}, and using the Taylor
approximation described above, we obtain the following
approximate variational objective function,
up to an additive constant independent of the model and
variational parameters.
The P-IG terms are combined with the entropy of
$q(\omega_{ij})$ through the normalizing constant
$Z_{\widetilde c_{ij}}$:
\begin{align*}
\label{eq:zidm_elbo}
\mathcal L_{\mathrm{ZIDM}}
={}&
\frac{1}{2}
\sum_{i=1}^n
\left\{
\log|\mathbf A_i|
-
\operatorname{tr}(\mathbf A_i)
-
\mathbf a_i^\top\mathbf a_i
-
2N_i\log S_i
\right\}
\nonumber\\
&+
\sum_{i=1}^n
\sum_{j=1}^p
\Bigg\{
(1-\widetilde{\pi}_{ij})
\left[
\log(1-\zeta_j)
-
\log(1-\widetilde{\pi}_{ij})
\right]
\nonumber\\
&\hspace{27mm}
+
\widetilde{\pi}_{ij}
\left[
\log\zeta_j
-
\log\widetilde{\pi}_{ij}
\right]
\Bigg\}
\nonumber\\
&+
\sum_{i=1}^n
\sum_{j=1}^p
(1-\widetilde{\pi}_{ij})
\Bigg[
\log\Gamma
\left(
\widetilde{\theta}_{ij}+y_{ij}
\right)
-
\left(
\widetilde{\theta}_{ij}+y_{ij}
\right)
\log
\left(
1+\frac{N_i}{S_i}
\right)
\nonumber\\
&\hspace{27mm}
+
\gamma_{\mathrm E}\widetilde{\theta}_{ij}
-
\gamma_{\mathrm E}\widetilde c_{ij}
-
\boldsymbol{\gamma}_j^\top
\mathbf A_i
\boldsymbol{\gamma}_j
-
\log\Gamma(\widetilde c_{ij})
\Bigg].
\end{align*}

For computation, the approximate variational objective for ZIDM-LVM
are implemented in \texttt{TMB} \citep{tmb}, where automatic differentiation is
used to evaluate the objective functions and their gradients for
numerical optimization. The model and variational parameters are
optimized jointly, including the auxiliary Gamma variational rate
parameters $S_i$.

For observed zeros, the variational probabilities
$\widetilde{\pi}_{ij}$ are retained directly in the variational
objective. To improve computational efficiency, following the variational
classification strategy of \citet{Zeng2023}, we additionally
consider a hard classification step in which
$\widetilde{\pi}_{ij}$ is replaced by the binary assignment
$\widetilde z_{ij}$ in subsequent parameter updates, defined as
\begin{equation*}
\label{eq:hard_classification}
\widetilde{\pi}_{ij}
\;\longrightarrow\;
\widetilde z_{ij}
=
\begin{cases}
0,
& y_{ij}>0,\\[3pt]
\mathbf 1\{\widetilde{\pi}_{ij}\geq\tau\},
& y_{ij}=0,
\end{cases}
\end{equation*}
where $\tau\in(0,1)$ is a classification threshold, with
$\tau=0.5$ used by default. Thus, an observed zero with
$\widetilde{\pi}_{ij}<0.5$ is assigned to the sampling component
($\widetilde z_{ij}=0$), whereas one with
$\widetilde{\pi}_{ij}\geq0.5$ is assigned to the structural-zero
component ($\widetilde z_{ij}=1$).

The binary assignments remove the need to propagate uncertainty in
$z_{ij}$ through subsequent parameter updates. In addition, the
Bernoulli entropy terms
$-\widetilde{\pi}_{ij}\log\widetilde{\pi}_{ij}
-(1-\widetilde{\pi}_{ij})\log(1-\widetilde{\pi}_{ij})$
are no longer required once the assignments are fixed. Together,
these simplifications reduce the computational burden of model
fitting, which is particularly useful for high-dimensional
microbiome data. Thus, we consider two approaches: classification
VA, which uses hard assignments for the structural-zero indicators,
and conventional VA, which retains the variational probabilities
$\widetilde{\pi}_{ij}$ throughout the optimization process.

\section{Simulation design}
\label{sec:simulation}
\subsection{Data generating mechanism}

We considered three fully parametric data-generating mechanisms
based on the zero-inflated negative binomial (ZINB-LVM),
ZIMN-LVM, and 
ZIDM-LVM distributions within the
GLLVM framework. The ZINB-LVM and ZIDM-LVM data-generating mechanisms are described
in the main text, while the ZIMN-LVM data-generating mechanism and its
corresponding simulation settings are provided in
Appendix~\ref{app:zimn_dgm}. The simulated datasets were generated
using parameter values drawn from prespecified distributions as shown in Table~\ref{tab:simulation_parameters}.

For both the ZINB-LVM and ZIDM-LVM, the latent variable
$\mathbf{u}_i$ is independently generated from
$N_d(\mathbf{0},\mathbf{I}_d)$ for each sample $i$, while the
structural-zero indicator $z_{ij}$ is independently generated from
$\mathrm{Bernoulli}(\zeta_j)$, where $z_{ij}=1$ indicates a
structural zero.
\textcolor{black}{
\paragraph{ZINB-LVM data-generating mechanism.}
The mean parameter of the negative binomial distribution is specified as
\[
\log(\mu_{ij})
=
\alpha_i+\beta_{0j}
+\mathbf{x}_i^\top\boldsymbol{\beta}_j
+\mathbf{u}_i^\top\boldsymbol{\gamma}_j.
\]
Conditional on $\mu_{ij}$ and $z_{ij}$, the observed count $Y_{ij}$
is set to zero when $z_{ij}=1$; otherwise, it is independently
generated from $\mathrm{NB}(\mu_{ij},\kappa)$, where $\kappa$ denotes
the negative binomial dispersion parameter.
}

\textcolor{black}{
\paragraph{ZIDM-LVM data-generating mechanism.}
The linear predictor and the corresponding concentration parameter
are defined as
\[
\eta_{ij}
=
\beta_{0j}
+\mathbf{x}_i^\top\boldsymbol{\beta}_j
+\mathbf{u}_i^\top\boldsymbol{\gamma}_j,
\qquad
\theta_{ij}=\exp(\eta_{ij}).
\]
The latent abundance parameter is then generated from
$\lambda_{ij}\sim\mathrm{Gamma}(\theta_{ij},1)$.
The observed count vector $\mathbf{Y}_i$ is generated according to
\[
\mathbf{Y}_i
\sim
\begin{cases}
\delta_{\mathbf{0}_p}(\cdot),
& z_{ij}=1,\ \forall j,\\[1mm]
\mathrm{Multinomial}(N_i,\rho_{i1}^{0},\ldots,\rho_{ip}^{0}),
& \text{otherwise},
\end{cases}
\]
where $\lambda_{ij}^{0}=(1-z_{ij})\lambda_{ij}$ and
$\rho_{ij}^{0}=\lambda_{ij}^{0}/\sum_{\ell=1}^{p}\lambda_{i\ell}^{0}$.
}

\paragraph{Parameter setting}

The common parameter settings for the   data-generating mechanisms are summarized in Table~\ref{tab:simulation_parameters}. For all  mechanisms, the regression coefficients and latent loadings were generated from the same distributions.
In the simulation study, we considered both high-dimensional and
low-dimensional settings, namely $(n,p)=(50,100)$ and $(100,50)$,
respectively, where $n$ denotes the sample size and $p$ denotes
the number of taxa.

\begin{table}[htbp]
\centering
\caption{Common parameter settings for the ZINB-LVM and ZIDM-LVM data-generating mechanisms.}
\label{tab:simulation_parameters}
\small
\renewcommand{\arraystretch}{1.20}
\setlength{\tabcolsep}{8pt}

\begin{tabular}{lcc}
\toprule
\textbf{Parameter}
&
\textbf{ZINB}
&
\textbf{ZIDM}
\\
\midrule

$\boldsymbol{\beta}_0$
&
\multicolumn{2}{c}{$\boldsymbol{\beta}_0 \sim \mathcal{U}(-1,1)$}
\\

$\mathbf{x}_{\cdot1}$
&
\multicolumn{2}{c}{$\mathbf{x}_{\cdot1}\sim\mathcal{N}(0,1)$}
\\

$\mathbf{x}_{\cdot2}$
&
\multicolumn{2}{c}{$\mathbf{x}_{\cdot2}\sim\mathrm{Exp}(1)$}
\\

$\boldsymbol{\beta}_1,\boldsymbol{\beta}_2$
&
\multicolumn{2}{c}{$
\boldsymbol{\beta}_1,\boldsymbol{\beta}_2
\sim\mathcal{U}(-1,1)$}
\\

$\boldsymbol{\gamma}_j$
&
\multicolumn{2}{c}{$
\boldsymbol{\gamma}_j\sim\mathcal{U}(-2,2)$}
\\

$\mathbf{u}_i$
&
\multicolumn{2}{c}{$
\mathbf{u}_i\sim\mathcal{N}(\mathbf0,\mathbf I_d)$}
\\

Overdispersion $\kappa$
&
$1$
&
--
\\

\bottomrule
\end{tabular}
\end{table}

We jointly calibrated the structural zero-inflation level and sampling
intensity so that the mean observed-zero proportion was approximately
50--55\%. Three levels of structural zero inflation were considered:
low, medium, and high, corresponding to
$\zeta_j\sim\mathcal{U}(0.05,0.15)$,
$\mathcal{U}(0.20,0.30)$, and
$\mathcal{U}(0.35,0.45)$, respectively.
The sampling intensity was calibrated separately for each
data-generating mechanism, dimension, and zero-inflation level.
Candidate sequencing-depth distributions $N_i$ for ZIDM-LVM and
candidate values of $\alpha_i$ for ZINB-LVM were evaluated using
independently generated datasets.The resulting settings for
ZIDM-LVM and ZINB-LVM are shown in Table~\ref{tab:simulation_sparsity}. 
\textcolor{black}{The sample-specific intercept ($\alpha_i$) was included only in the ZINB-LVM to control the overall count intensity. For the ZIDM-LVM, no additional sample-specific intercept was included in the data-generating mechanisms, and the total count was determined by the sequencing depth ($N_i$).}

\begin{table}[htbp]
\centering
\caption{
Calibrated zero-inflation and sampling-intensity settings for ZIDM-LVM and
ZINB-LVM. Here, $\zeta_j$ is the taxon-specific structural-zero probability,
$\mathcal{U}_d(a,b)$ denotes the discrete uniform distribution on
$[a,b]$, $N_i$ is the ZIDM-LVM sequencing depth, and $\alpha_i$
is the ZINB-LVM row effect controlling count intensity.
}
\label{tab:simulation_sparsity}
\small
\renewcommand{\arraystretch}{1.20}
\setlength{\tabcolsep}{5pt}

\begin{tabular}{lc|cc|cc}
\toprule
\multicolumn{2}{c|}{\textbf{Zero inflation}}
&
\multicolumn{2}{c|}{$(n,p)=(50,100)$}
&
\multicolumn{2}{c}{$(n,p)=(100,50)$}
\\
\cmidrule(lr){1-2}
\cmidrule(lr){3-4}
\cmidrule(lr){5-6}

\textbf{ZI level}
&
\textbf{$\zeta_j$}
&
\textbf{ZINB $\alpha_i$}
&
\textbf{ZIDM $N_i$}
&
\textbf{ZINB $\alpha_i$}
&
\textbf{ZIDM $N_i$}
\\
\midrule

Low
&
$\mathcal{U}(0.05,0.15)$
&
$0.25$
&
$\mathcal{U}_d(320,480)$
&
$0.25$
&
$\mathcal{U}_d(140,210)$
\\

Medium
&
$\mathcal{U}(0.20,0.30)$
&
$1$
&
$\mathcal{U}_d(800,1200)$
&
$1$
&
$\mathcal{U}_d(320,480)$
\\

High
&
$\mathcal{U}(0.35,0.45)$
&
$2.25$
&
$\mathcal{U}_d(10000,15000)$
&
$2$
&
$\mathcal{U}_d(4000,6000)$
\\

\bottomrule
\end{tabular}

\end{table}

We compared the proposed methods with several existing approaches:
(a) ZINB-LVM with sample-specific row
effects $\alpha_i$, fitted using the \texttt{gllvm} package
\citep{Niku2019gllvm} (ZINB);
(b) the proposed ZIMN-LVM (ZIMN);
(c) the proposed ZIDM-LVM
(ZIDM); \textcolor{black}{For both the ZIMN-LVM and ZIDM-LVM, no additional
sample-specific intercept $\alpha_i$ was included
in the linear predictor, and model estimation was
performed in \texttt{TMB}.}
(d) the Bayesian ZIDM model of \citet{Koslovsky2023} (ZIDM-K), fitted using the accompanying \texttt{ZIDM} package. \textcolor{black}{Following the MCMC settings of \citet{Koslovsky2023},
we ran 20,000 MCMC iterations and discarded the first
50\% as burn-in.}
(e) zero replacement (ZR), in which zero counts were replaced by 0.5
before renormalizing each sample.
For brevity, the abbreviations in parentheses are used to denote
the corresponding methods in the figures and tables. \textcolor{black}{For each simulation setting, 1500 independent datasets were
generated. Model fits that failed to converge or did not
produce valid estimates were excluded from the calculation
of the corresponding performance measures.
The proportion of valid results was close to 100\% for
all methods except ZIMN-LVM. Under the ZINB-LVM generator,
this proportion was as low as 87.7\% for ZIMN-LVM
in some simulation settings.
}

For the proposed methods (ZIDM-LVM and ZIMN-LVM), the main analysis
used conventional VA, retaining the estimated variational
ZI probabilities $\widetilde{\pi}_{ij}$ throughout
optimization. We additionally examined classification VA, which
uses hard assignments for the structural-zero indicators, in a
sensitivity analysis in Appendix~\ref{app:soft_hard}. Its recovery
performance relative to conventional VA was evaluated using the
ratio of classification VA to conventional VA recovery errors.

All computations were performed on the Roihu supercomputer in Finland. All code used in this study is publicly available at
\url{https://github.com/tangwenq/ZIDM-LVM-simulation-analysis}.

\subsection{Comparison of underlying composition}

Following \citet{Zeng2023}, we first compared the underlying
composition with the estimated composition.
The underlying composition $\rho_{ij}$ was obtained directly from the
data-generating process before applying the structural-zero indicators
$z_{ij}$.
\textcolor{black}{
Under the ZINB-LVM, the underlying composition is obtained by
normalizing the exponentiated linear predictors as
$\rho_{ij}^{\mathrm{ZINB}}
=
\frac{\mu_{ij}}{\sum_{\ell=1}^{p}\mu_{i\ell}}$.
Under the ZIDM-LVM, the underlying composition is obtained by
normalizing the latent abundance parameters as
$\rho_{ij}^{\mathrm{ZIDM}}
=
\frac{\lambda_{ij}}{\sum_{\ell=1}^{p}\lambda_{i\ell}}$.}

For ZINB-LVM, ZIMN-LVM, and ZIDM-LVM, the estimated composition was
calculated using the quadratic correction introduced previously to
account for variational uncertainty in the latent scores:
\begin{equation*}
\hat{\rho}_{ij}
=
\frac{
\exp\!\left(
\hat{\eta}_{ij}
+
\frac{1}{2}
\hat{\boldsymbol{\gamma}}_j^\top
\hat{\mathbf A}_i
\hat{\boldsymbol{\gamma}}_j
\right)
}{
\sum_{l=1}^{p}
\exp\!\left(
\hat{\eta}_{il}
+
\frac{1}{2}
\hat{\boldsymbol{\gamma}}_l^\top
\hat{\mathbf A}_i
\hat{\boldsymbol{\gamma}}_l
\right)
}.
\label{eq:rho_hat}
\end{equation*}

For ZIDM-K, the estimated composition was obtained by averaging the
baseline compositions over the retained posterior MCMC draws, whereas
for ZR, zero counts were replaced by 0.5 and each sample was
renormalized to sum to one.

The estimated compositions were compared with the underlying compositions
using four evaluation metrics, following the composition-recovery
evaluation in \citet{Zeng2023}. The metrics are the  Frobenius norm error
\[
\sqrt{
\sum_{i=1}^{n}\sum_{j=1}^{p}
(\hat{\rho}_{ij}-\rho_{ij})^2
},
\]
the average Kullback--Leibler (KL) divergence
\[
\frac{1}{n}
\sum_{i=1}^{n}\sum_{j=1}^{p}
\rho_{ij}
\log\left(
\frac{\rho_{ij}}{\hat{\rho}_{ij}}
\right),
\]
the Shannon's index mean squared error
\[
\frac{1}{n}
\sum_{i=1}^{n}
\left(
\sum_{j=1}^{p}
\hat{\rho}_{ij}\log(\hat{\rho}_{ij})
-
\sum_{j=1}^{p}
\rho_{ij}\log(\rho_{ij})
\right)^2,
\]
and the Simpson's index mean squared error
\[
\frac{1}{n}
\sum_{i=1}^{n}
\left(
\sum_{j=1}^{p}\rho_{ij}^{2}
-
\sum_{j=1}^{p}\hat{\rho}_{ij}^{2}
\right)^2.
\]

\begin{figure}[!htbp]
    \centering

    \begin{subfigure}{\textwidth}
        \centering
        \includegraphics[
            width=\linewidth,
            height=0.38\textheight,
            keepaspectratio
        ]{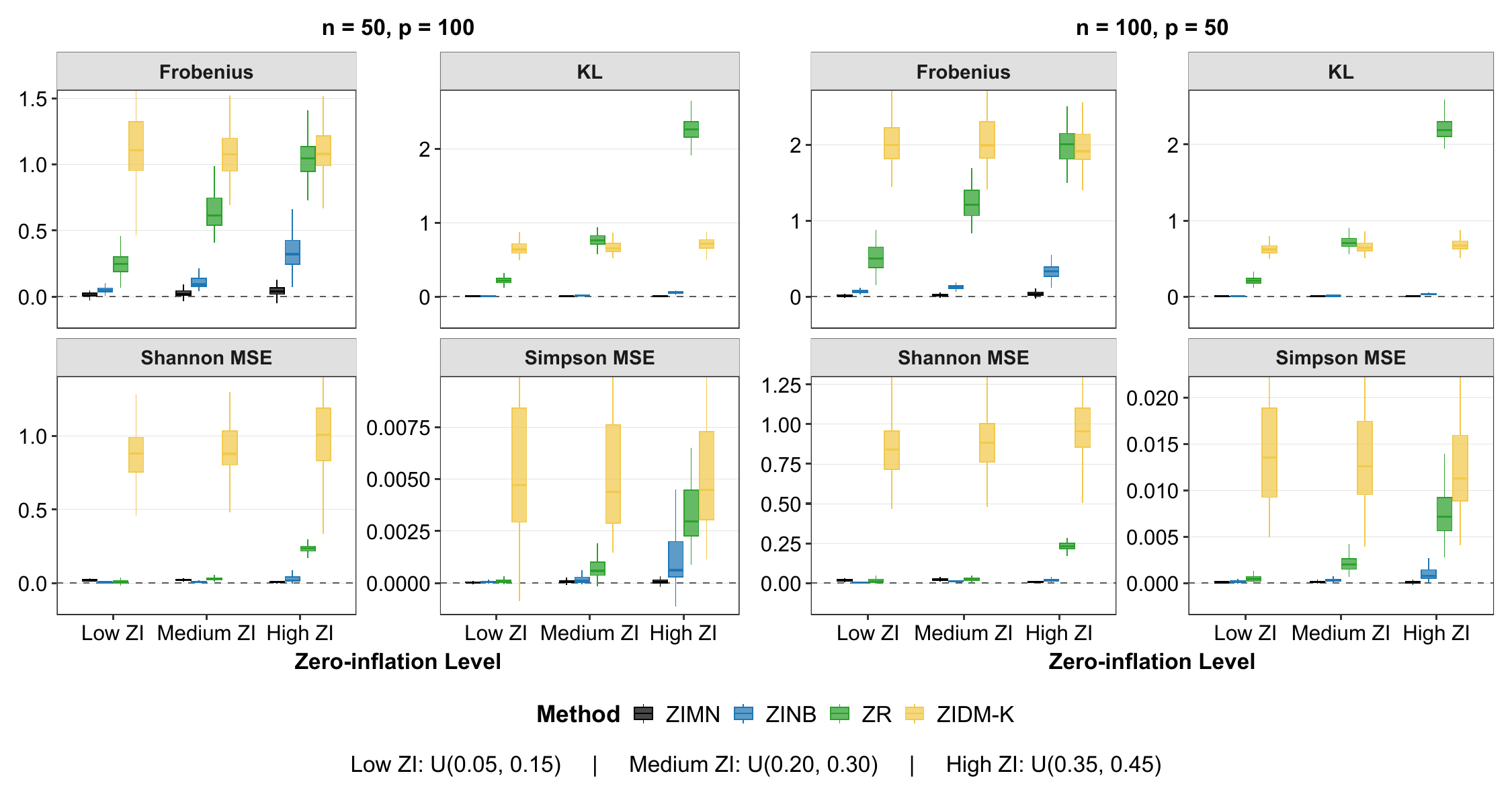}
        \caption{ZIDM-LVM}
        \label{fig:zidm_composition}
    \end{subfigure}

    \vspace{4mm}

    \begin{subfigure}{\textwidth}
        \centering
        \includegraphics[
            width=\linewidth,
            height=0.38\textheight,
            keepaspectratio
        ]{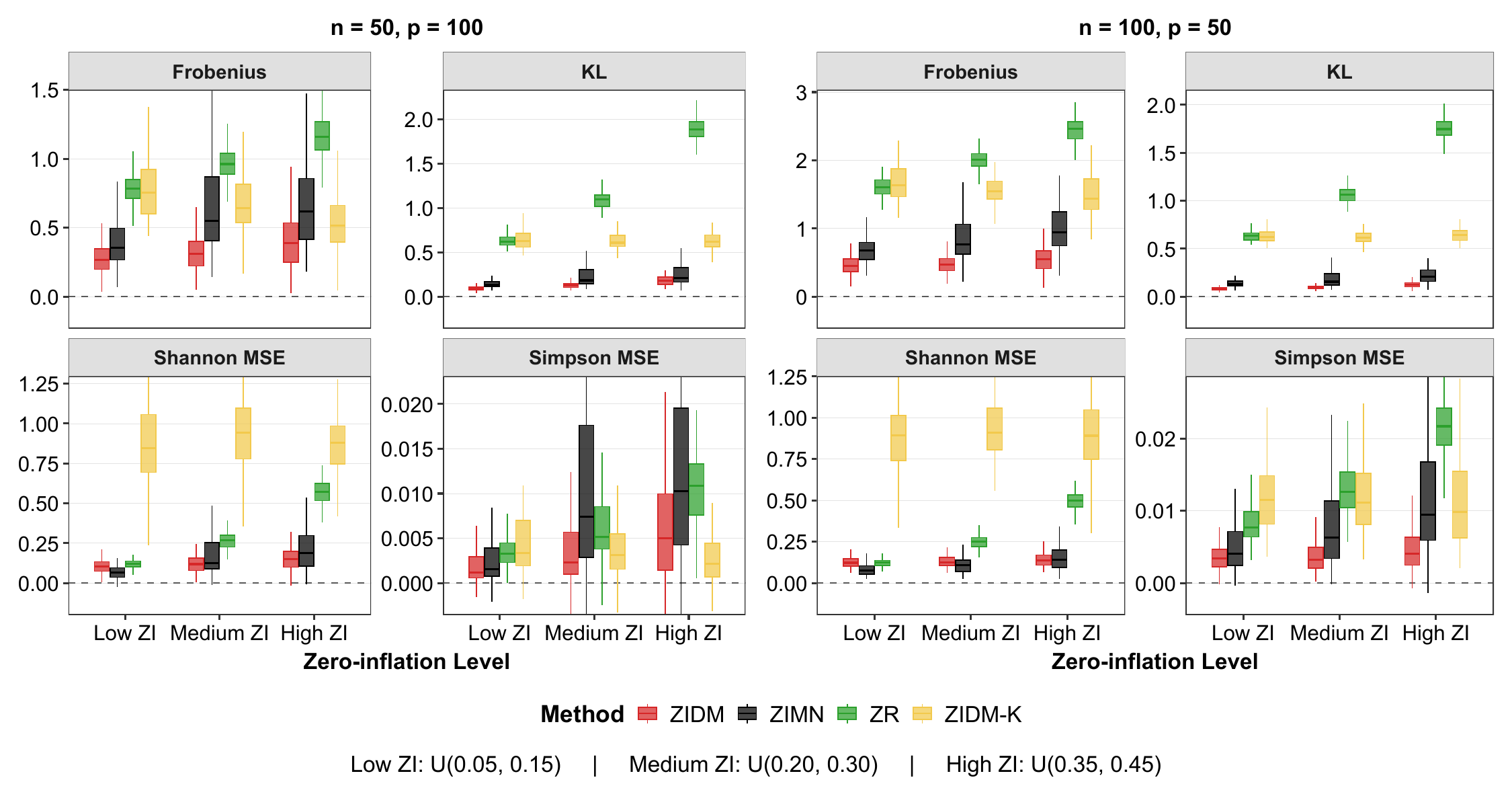}
        \caption{ZINB-LVM}
        \label{fig:zinb_composition}
    \end{subfigure}

\caption{Differences in composition recovery errors relative
to the corresponding generating model for (a) ZIDM-LVM and (b) ZINB-LVM data
across three levels of structural zero inflation and two
dimensional settings. Positive values indicate larger
recovery errors.}
    \label{fig:composition_recovery}
\end{figure}

Figures~\ref{fig:zidm_composition} and
\ref{fig:zinb_composition} show the differences in composition
recovery errors relative to the corresponding generating model,
with positive values indicating larger errors. Both figures
show similar trends, with error differences generally increasing
as the proportion of structural zeros increased. These increases
were relatively small for the other model-based methods, whereas
ZR and ZIDM-K generally exhibited larger error differences,
greater variability, and more pronounced increases.

For data generated from ZIDM-LVM
(Figure~\ref{fig:zidm_composition}), ZIMN-LVM and ZINB-LVM generally
showed relatively small error differences compared with ZIDM-LVM,
indicating competitive composition recovery. For data generated from ZINB-LVM
(Figure~\ref{fig:zinb_composition}), ZIDM-LVM generally showed
smaller error differences than ZIMN-LVM, although both methods
exhibited larger errors than the generating model.
Similar trends were observed for data generated from ZIMN-LVM,
as shown in Figure~\ref{fig:zimn_composition} in the Appendix~\ref{app:zimn_dgm}.

In comparison, the differences between methods and the
variability in recovery error differences were larger
for data generated from ZINB-LVM than for data generated from ZIDM-LVM.

Table~\ref{tab:runtime_zidm_zinb} summarizes the mean model fitting times over 100 simulation replicates for each setting under the ZIDM-LVM and ZINB-LVM data-generating mechanisms. Overall, ZIMN-LVM was generally the fastest method, followed by ZIDM-LVM, whereas ZINB-LVM was typically more computationally intensive.
\textcolor{black}{ZIDM-K exhibited relatively stable mean fitting times across simulation settings, ranging from approximately 54 to 63 seconds. The reported times represent model fitting times only and exclude the computation of confidence and credible intervals.}

\begin{table}[!htbp]
\centering
\caption{Mean model fitting time (seconds) over 100 simulation
replicates under the ZIDM-LVM and ZINB-LVM data-generating
mechanisms, for $(n,p)=(50,100)$ and $(100,50)$.
}
\label{tab:runtime_zidm_zinb}

\scriptsize
\setlength{\tabcolsep}{4.2pt}
\renewcommand{\arraystretch}{1.12}

\makebox[\textwidth][c]{%
\begin{tabular}{lcccc@{\hspace{8mm}}cccc}
\toprule

& \multicolumn{4}{c}{$(n,p)=(50,100)$}
& \multicolumn{4}{c}{$(n,p)=(100,50)$} \\

\cmidrule(lr){2-5}
\cmidrule(lr){6-9}

& ZIDM & ZIMN & ZINB & ZIDM-K
& ZIDM & ZIMN & ZINB & ZIDM-K \\

\midrule

\multicolumn{9}{c}{ZIDM-LVM generator} \\

Low
& 12.03 & 2.96 & 20.26 & 59.74
& 10.33 & 5.06 & 22.02 & 62.83 \\

Medium
& 8.63 & 2.85 & 24.27 & 59.17
& 6.67 & 3.46 & 19.03 & 56.03 \\

High
& 5.62 & 4.61 & 25.41 & 60.83
& 5.54 & 5.92 & 19.78 & 54.01 \\

\midrule

\multicolumn{9}{c}{ZINB-LVM generator} \\

Low
& 9.23 & 6.19 & 43.10 & 58.59
& 14.85 & 21.55 & 36.99 & 58.60 \\

Medium
& 12.40 & 9.03 & 54.20 & 59.96
& 6.50 & 6.31 & 35.90 & 59.70 \\

High
& 14.91 & 16.79 & 46.74 & 58.63
& 7.66 & 6.65 & 30.69 & 55.31 \\

\bottomrule
\end{tabular}%
}

\end{table}
\FloatBarrier
\subsection{Comparison of ordination recovery}
\label{subsec:ordination_recovery}

Following the approach of \citet{Tang2025}, we evaluated
ordination recovery by comparing the true and estimated
ordinations for the GLLVM-based methods. Since latent
ordinations are identifiable only up to orientation and scale,
we used Procrustes analysis \citep{HurleyCattell1962}, which
finds the optimal alignment between two ordinations through
translation, rotation, reflection, and uniform scaling.
Ordination recovery was evaluated by the Procrustes error
\begin{equation*}
D_{\mathrm{Proc}}
=
\sum_{i=1}^{n}
\sum_{r=1}^{d}
\left(
\hat u_{ir,\mathrm{fitted}}
-
u_{ir,\mathrm{true}}
\right)^2,
\end{equation*}
where $\hat u_{ir,\mathrm{fitted}}$ denotes the estimated latent
score after Procrustes alignment, with each latent dimension
standardized to have zero mean and unit variance before
alignment. The analysis was implemented using the
\texttt{procrustes} function in the \texttt{vegan} package \citep{vegan13}.
Smaller values indicate better recovery of the true ordination
structure.

Table~\ref{tab:ordination_recovery} reports the mean Procrustes
errors. Across both generating models, the correctly specified
model generally achieved the smallest errors. Ordination recovery
also tended to become poorer as the proportion of structural
zeros increased, particularly for the other models.
ZIMN-LVM generally showed larger errors than ZIDM-LVM. In comparison,
the differences between methods were substantially larger
for data generated from ZINB-LVM than for data generated from ZIDM-LVM.

\begin{table}[!htbp]
\caption{Mean Procrustes errors for ordination recovery
for data generated from ZIDM-LVM and ZINB-LVM. Smaller values
indicate better recovery of the true ordination structure.}

\label{tab:ordination_recovery}

\scriptsize
\setlength{\tabcolsep}{5.0pt}
\renewcommand{\arraystretch}{1.12}

\makebox[\textwidth][c]{%
\begin{tabular}{lccc@{\hspace{8mm}}ccc}

\toprule

& \multicolumn{3}{c}{$(n,p)=(50,100)$}
& \multicolumn{3}{c}{$(n,p)=(100,50)$} \\

\cmidrule(lr){2-4}
\cmidrule(lr){5-7}

& ZIDM
& ZIMN
& ZINB
& ZIDM
& ZIMN
& ZINB \\

\midrule

\multicolumn{7}{c}{ZIDM-LVM generator} \\

Low
& \textbf{5.57}
& 5.75
& 5.83
& \textbf{9.19}
& 9.81
& 10.29 \\

Medium
& \textbf{5.52}
& 5.76
& 6.02
& \textbf{8.67}
& 9.35
& 10.41 \\

High
& \textbf{5.54}
& 5.89
& 6.95
& \textbf{9.18}
& 9.18
& 13.37 \\

\midrule

\multicolumn{7}{c}{ZINB-LVM generator} \\

Low
& 14.38
& 19.70
& \textbf{7.03}
& 30.84
& 44.82
& \textbf{13.80} \\

Medium
& 16.66
& 20.81
& \textbf{7.36}
& 35.09
& 52.52
& \textbf{14.60} \\

High
& 20.36
& 19.69
& \textbf{7.55}
& 40.95
& 57.91
& \textbf{16.24} \\

\bottomrule

\end{tabular}%
}

\end{table}

\FloatBarrier

\subsection{Comparison of parameters}
\begin{figure}[!htbp]
    \centering

    \begin{subfigure}{\textwidth}
        \centering
        \includegraphics[
            width=1\textwidth
        ]{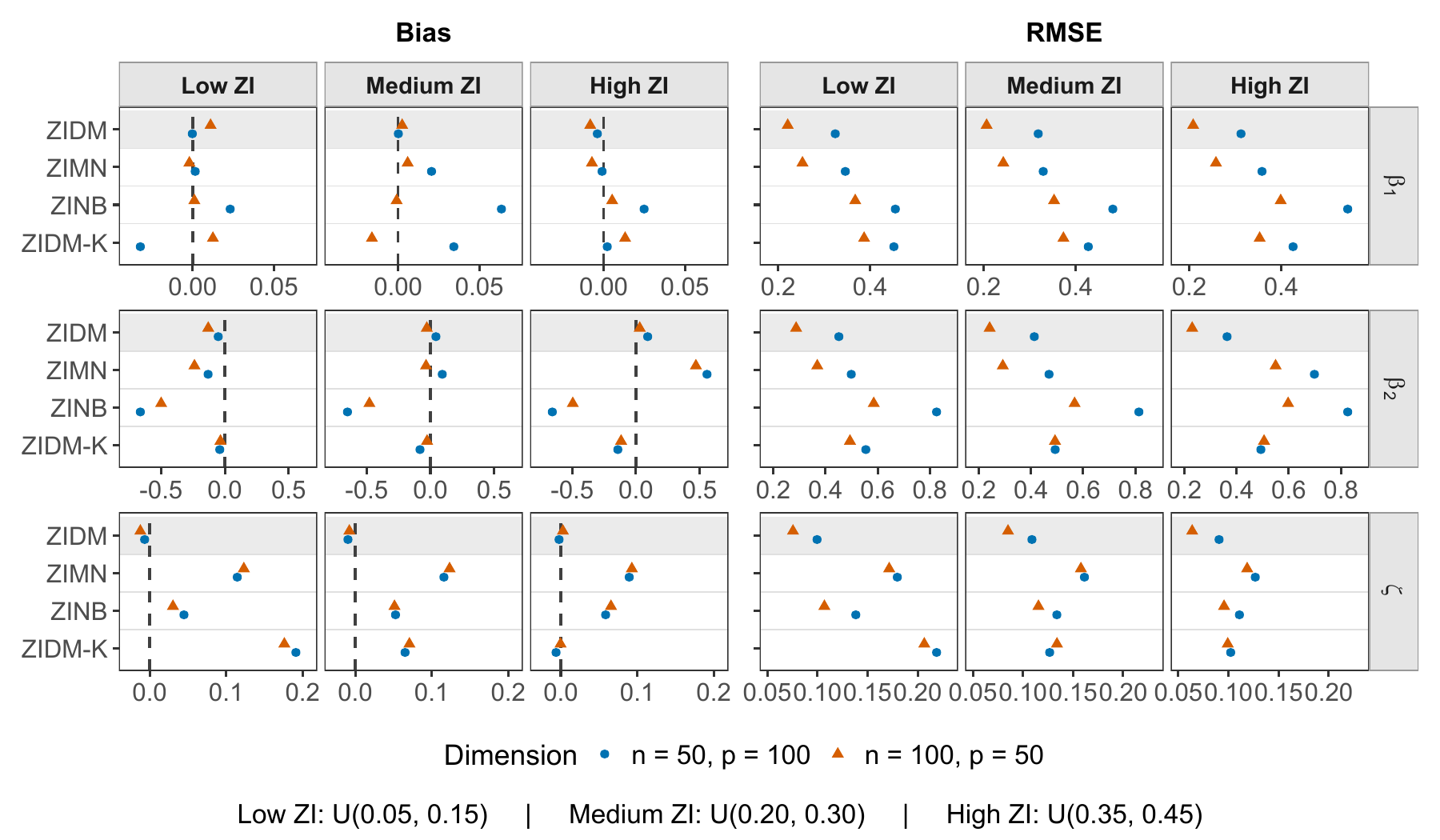}
        \caption{ZIDM-LVM}
        \label{fig:parameter_zidm}
    \end{subfigure}

    \vspace{3mm}

    \begin{subfigure}{\textwidth}
        \centering
        \includegraphics[
            width=1\textwidth
        ]{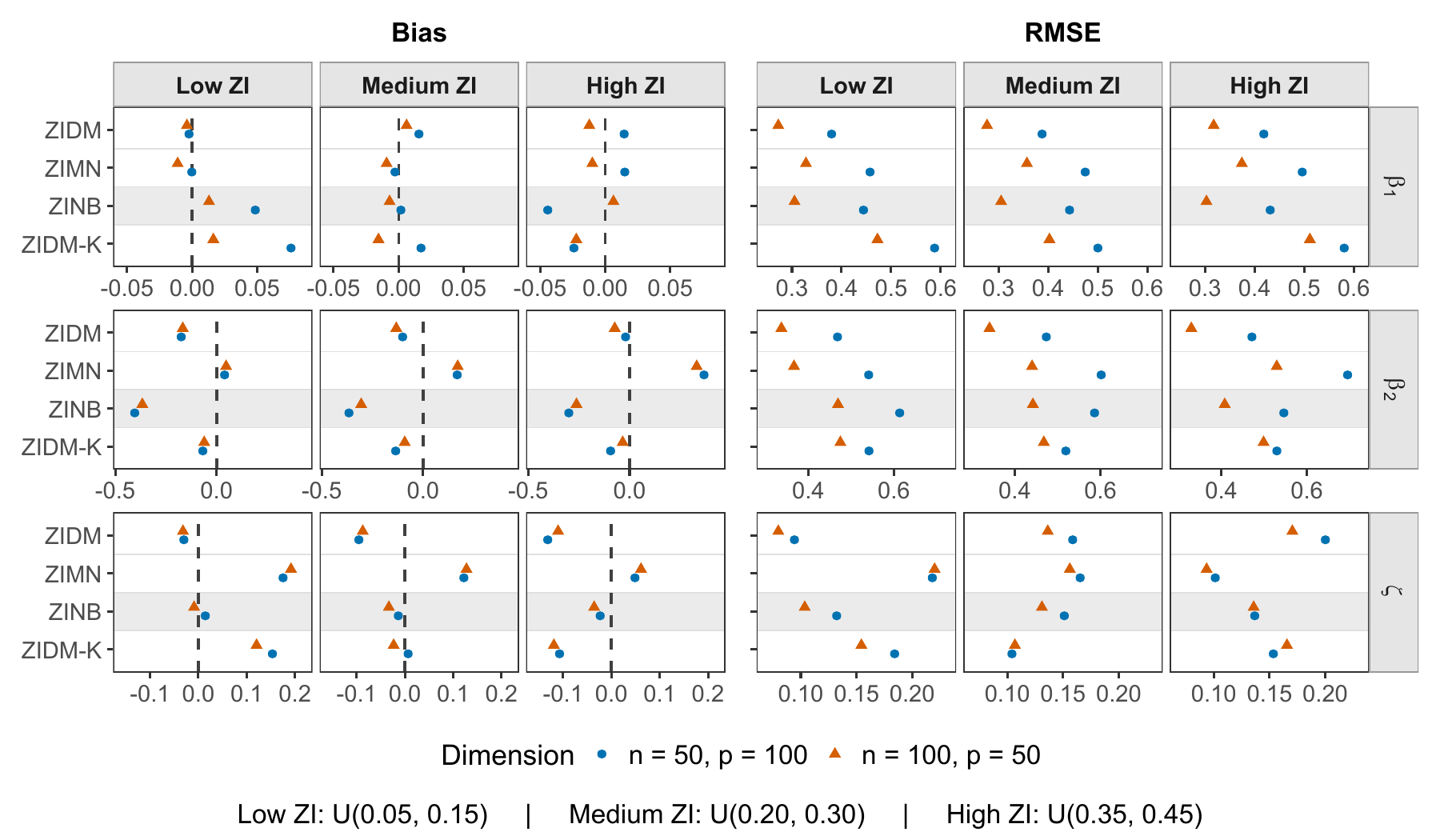}
        \caption{ZINB-LVM}
        \label{fig:parameter_zinb}
    \end{subfigure}

    \caption{Mean bias and RMSE of parameter estimates for data
generated from (a) ZIDM-LVM and (b) ZINB-LVM across three levels
of structural zero inflation and two dimensional settings.
Gray shading highlights the method corresponding to the
generating model. The dashed line indicates zero bias.}
\label{fig:parameter_recovery}
\end{figure}

Parameter estimation accuracy was evaluated for the environmental
coefficients $\boldsymbol{\beta}_{j}$ and zero-inflation probabilities
$\zeta_j$.
Let $\psi$ denote a generic true parameter and
$\hat{\psi}^{(r)}$ its estimate from simulation replicate
$r=1,\ldots,R$. Estimation performance was summarized using
Bias and Root Mean Squared Error (RMSE):

\[
\operatorname{Bias}(\hat{\psi})
=
\frac{1}{R}
\sum_{r=1}^{R}
\left(
\hat{\psi}^{(r)}-\psi
\right),
\]

and

\[
\operatorname{RMSE}(\hat{\psi})
=
\sqrt{
\frac{1}{R}
\sum_{r=1}^{R}
\left(
\hat{\psi}^{(r)}-\psi
\right)^2
}.
\]
Figure~\ref{fig:parameter_recovery} summarizes the parameter
estimation performance in terms means of bias and RMSE.

We additionally use the mean confidence interval (CI) width and
empirical coverage to assess regression coefficients. \textcolor{black}{For the frequentist methods, 95\% Wald confidence intervals
were constructed using the estimated regression coefficients
and their standard errors. For the ZIDM-LVM and ZIMN-LVM,
standard errors were derived from the covariance matrix
approximated by the inverse of the Schur complement of the
Hessian matrix computed using TMB, while standard errors
and confidence intervals for the ZINB-LVM were obtained
directly from the \texttt{gllvm} package.
For the Bayesian ZIDM method, 95\% credible intervals
were obtained from the 2.5th and 97.5th posterior quantiles}. 

As seen in Table~\ref{tab:ci_recovery}, for the regression coefficients, ZIDM-LVM generally showed
small biases and relatively low RMSEs across both generating
models. For data generated from ZIDM-LVM, ZIDM-LVM generally
achieved the lowest RMSEs and coverage probabilities closest
to the nominal level of 0.95\textcolor{black}{, ranging from 0.89 to 0.94}.
For data generated from ZINB-LVM, ZIDM-LVM remained competitive
with ZINB-LVM in terms of bias and RMSE, although its coverage
probabilities were generally lower\textcolor{black}{, ranging from 0.72 to 0.86}.
\textcolor{black}{
Across both generating models, coverage probabilities were generally lower for $\beta_2$
associated with the exponentially distributed covariate, than for $\beta_1$
even for ZINB-LVM under its own generating model.
Confidence intervals that could not be computed due to
numerical difficulties in standard error estimation
were excluded from the calculation of empirical coverage
probabilities and mean interval widths, as reported in
Table~\ref{tab:ci_recovery}.}

For the zero-inflation parameter $\zeta$, ZIDM-LVM generally
showed small biases and relatively low RMSEs across both
generating models. ZINB-LVM also showed relatively small errors
in several settings, whereas ZIMN-LVM generally exhibited larger
positive biases and RMSEs. 

\textcolor{black}{
Overall, ZIDM-LVM maintained relatively stable point estimation
accuracy for both regression coefficients and zero-inflation
parameters across the two generating models, and interval
estimation was more sensitive to the data-generating conditions.
}

\begin{table}[!htbp]
\centering
\caption{Mean confidence interval (CI) width and empirical
coverage probability for the regression coefficient estimates
for data generated from ZIDM-LVM and ZINB-LVM. Values are reported
as CI width / \textbf{coverage}.}
\label{tab:ci_recovery}

\scriptsize
\renewcommand{\arraystretch}{1.15}
\setlength{\tabcolsep}{1.5pt}

\begin{tabular}{@{}l@{\hspace{3pt}}l
                *{4}{c}@{\hspace{3mm}}*{4}{c}@{}}
\toprule
& &
\multicolumn{4}{c}{$(n,p)=(50,100)$} &
\multicolumn{4}{c}{$(n,p)=(100,50)$} \\
\cmidrule(lr){3-6}
\cmidrule(lr){7-10}

ZI & Param.
& ZIDM & ZIMN & ZINB & ZIDM-K
& ZIDM & ZIMN & ZINB & ZIDM-K \\
\midrule

\multicolumn{10}{c}{ZIDM-LVM generator} \\
\midrule

Low
& $\beta_1$
& 1.22/\textbf{0.94}
& 1.08/\textbf{0.88}
& 1.60/\textbf{0.82}
& 0.89/\textbf{0.73}
& 0.84/\textbf{0.94}
& 0.72/\textbf{0.84}
& 1.64/\textbf{0.81}
& 0.62/\textbf{0.65} \\

& $\beta_2$
& 1.47/\textbf{0.94}
& 1.30/\textbf{0.86}
& 1.66/\textbf{0.53}
& 1.04/\textbf{0.61}
& 0.99/\textbf{0.92}
& 0.84/\textbf{0.69}
& 1.70/\textbf{0.47}
& 0.71/\textbf{0.48} \\

\addlinespace[2pt]

Medium
& $\beta_1$
& 1.16/\textbf{0.93}
& 0.93/\textbf{0.82}
& 1.27/\textbf{0.80}
& 0.89/\textbf{0.74}
& 0.81/\textbf{0.94}
& 0.64/\textbf{0.79}
& 1.34/\textbf{0.80}
& 0.63/\textbf{0.65} \\

& $\beta_2$
& 1.35/\textbf{0.93}
& 1.06/\textbf{0.80}
& 1.41/\textbf{0.52}
& 1.03/\textbf{0.63}
& 0.92/\textbf{0.94}
& 0.72/\textbf{0.77}
& 1.25/\textbf{0.46}
& 0.71/\textbf{0.51} \\

\addlinespace[2pt]

High
& $\beta_1$
& 1.13/\textbf{0.92}
& 0.85/\textbf{0.74}
& 1.39/\textbf{0.79}
& 0.95/\textbf{0.74}
& 0.76/\textbf{0.94}
& 0.59/\textbf{0.72}
& 0.96/\textbf{0.78}
& 0.65/\textbf{0.63} \\

& $\beta_2$
& 1.23/\textbf{0.89}
& 0.89/\textbf{0.33}
& 1.47/\textbf{0.54}
& 1.04/\textbf{0.67}
& 0.83/\textbf{0.92}
& 0.61/\textbf{0.28}
& 0.99/\textbf{0.49}
& 0.71/\textbf{0.55} \\

\midrule
\multicolumn{10}{c}{ZINB-LVM generator} \\
\midrule

Low
& $\beta_1$
& 1.13/\textbf{0.86}
& 1.06/\textbf{0.75}
& 1.45/\textbf{0.86}
& 0.87/\textbf{0.66}
& 0.74/\textbf{0.81}
& 0.70/\textbf{0.71}
& 1.34/\textbf{0.87}
& 0.59/\textbf{0.53} \\

& $\beta_2$
& 1.28/\textbf{0.82}
& 1.19/\textbf{0.75}
& 1.63/\textbf{0.76}
& 0.94/\textbf{0.59}
& 0.82/\textbf{0.72}
& 0.79/\textbf{0.71}
& 1.20/\textbf{0.65}
& 0.63/\textbf{0.49} \\

\addlinespace[2pt]

Medium
& $\beta_1$
& 1.13/\textbf{0.84}
& 1.01/\textbf{0.71}
& 1.41/\textbf{0.87}
& 0.87/\textbf{0.65}
& 0.74/\textbf{0.79}
& 0.72/\textbf{0.68}
& 1.01/\textbf{0.88}
& 0.61/\textbf{0.52} \\

& $\beta_2$
& 1.26/\textbf{0.82}
& 1.12/\textbf{0.66}
& 1.48/\textbf{0.79}
& 0.95/\textbf{0.60}
& 0.81/\textbf{0.74}
& 0.76/\textbf{0.60}
& 1.01/\textbf{0.70}
& 0.64/\textbf{0.49} \\

\addlinespace[2pt]

High
& $\beta_1$
& 1.15/\textbf{0.81}
& 0.98/\textbf{0.66}
& 1.38/\textbf{0.88}
& 0.91/\textbf{0.64}
& 0.77/\textbf{0.77}
& 0.70/\textbf{0.64}
& 1.02/\textbf{0.88}
& 0.64/\textbf{0.50} \\

& $\beta_2$
& 1.25/\textbf{0.79}
& 1.06/\textbf{0.50}
& 1.45/\textbf{0.80}
& 0.98/\textbf{0.63}
& 0.82/\textbf{0.75}
& 0.75/\textbf{0.44}
& 1.02/\textbf{0.73}
& 0.67/\textbf{0.49} \\

\bottomrule
\end{tabular}
\end{table}

\section{Real data analysis}
\label{sec:real_data_analysis}

We applied ZIDM-LVM, ZIMN-LVM, ZINB-LVM, and ZIPPCA
\citep{Zeng2023} to the arcto-alpine microbial data \citep{Kumar2017}
and the Nugent vaginal microbiome data \citep{ravel2011vaginal}
to compare their representations of microbial community
structure through two-dimensional ordinations.
ZIPPCA was fitted using the R package \texttt{ZIPPCAlnm}.
Centered log-ratio principal component analysis (CLR-PCA)
was additionally included as a reference method. \textcolor{black}{The comparison methods differed between the simulation study and the real-data analysis. ZIPPCA was excluded from the simulations because the implementation supports only one covariate, whereas the simulation settings included two. In contrast, ZIDM-K was included in the simulations but excluded from the real-data analysis because it does not provide latent ordination scores. Since the real-data analysis focused on two-dimensional ordination and count reconstruction, we used the same set of methods for both comparisons to maintain consistency.} 
The CLR transformation expresses each component as the
log-ratio to the geometric mean of the components within
the same sample.
A pseudocount of one was added to each count before
the transformation.
All models were fitted without sample covariates,
and we selected $k=2$ latent dimensions for the ordination
comparison.

\subsection{Data introduction}
\paragraph{Microbial data}
\textcolor{black}{
The arcto-alpine microbial community data originate from
\citet{Kumar2017} and comprise bacterial communities sampled
from three climatically distinct regions: Ny-\AA lesund in
the high Arctic, Kilpisj\"arvi in the low Arctic, and
Mayrhofen in the European Alps. These regions differ in climatic and environmental conditions,
providing an opportunity to investigate habitat-related
variation in microbial community composition.
We used the count data available as 
\texttt{microbialdata}
in the R package \texttt{gllvm} \citep{gllvmpackage}.
The dataset contains 56 soil samples and 985 bacterial OTUs,
with 63.3\% zero entries.
The counts are also strongly overdispersed, with the
variance exceeding the mean for most OTUs.
Sampling regions were used to examine the geographic
structure represented by the ordinations.}
\paragraph{Nugent data}

The vaginal microbiome data originate from the
cross-sectional study of North American women from four
self-reported ethnic groups by \citet{ravel2011vaginal},
in which bacterial communities were characterized by
pyrosequencing of barcoded 16S rRNA genes.
We used the processed count table and sample metadata
from MLRepo.
Following the prevalence criterion of
\citet{lubbe2021comparison}, we retained features detected
in at least 5\% of the samples.
The resulting count matrix contains 394 samples and
88 features, with 83.06\% zero entries.

\subsection{Ordination}

\begin{figure}[!htbp]
    \centering
    \makebox[\textwidth][c]{%
        \includegraphics[
            width=\textwidth
        ]{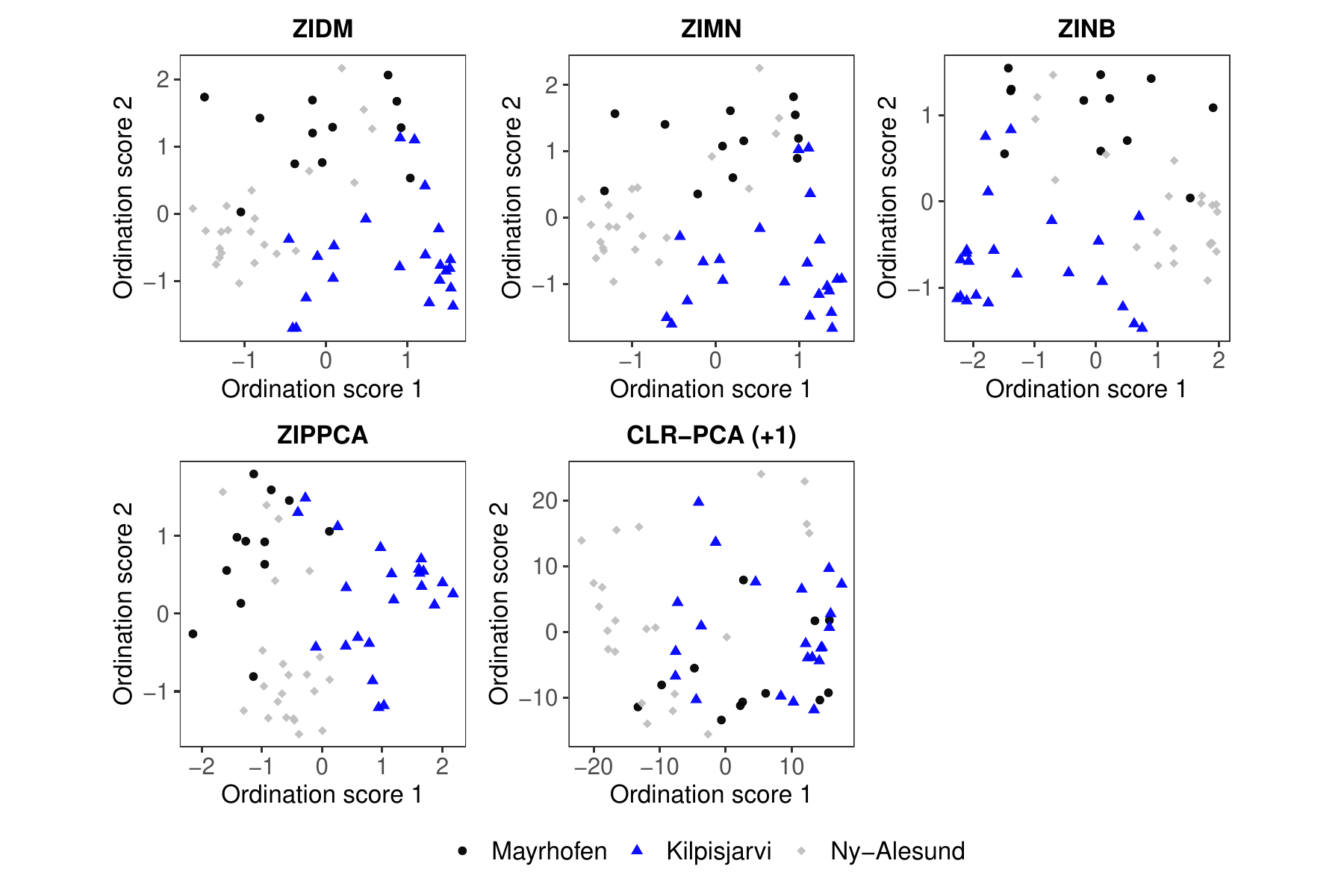}%
    }
    \caption{
        Two-dimensional ordinations of the microbial data
        obtained from ZIDM-LVM, ZIMN-LVM, ZINB-LVM, ZIPPCA, and CLR-PCA,
        colored by sampling region.
    }
    \label{fig:xnull_ordination}
\end{figure}

\begin{figure}[!htbp]
    \centering

    \begin{subfigure}{\textwidth}
        \centering
        \makebox[\linewidth][c]{%
            \includegraphics[width=\linewidth]
            {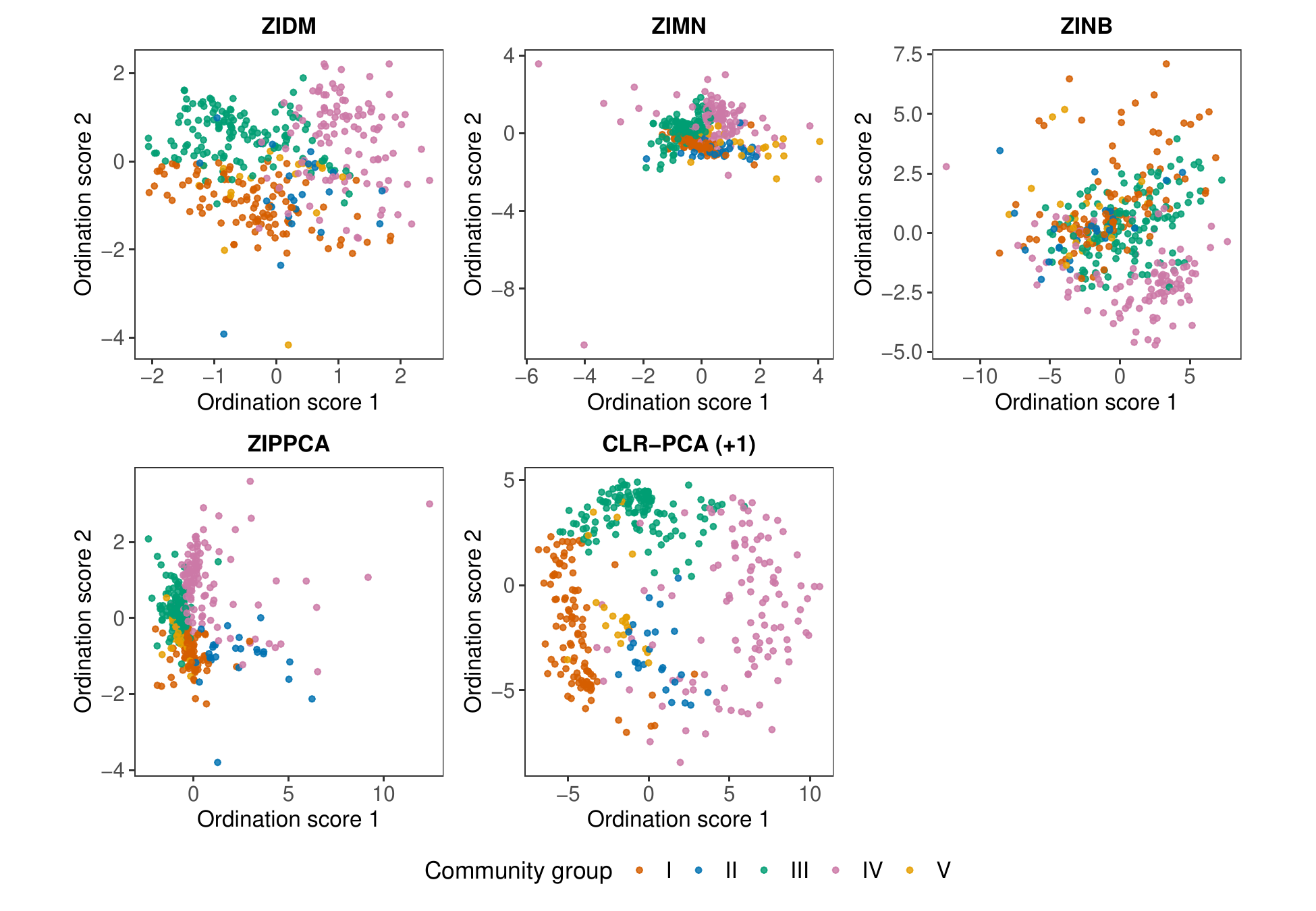}%
        }
        \caption{Community groups (I--V).}
        \label{fig:nugent_community}
    \end{subfigure}

    \vspace{0.5cm}

    \begin{subfigure}{\textwidth}
        \centering
        \makebox[\linewidth][c]{%
            \includegraphics[width=\linewidth]
            {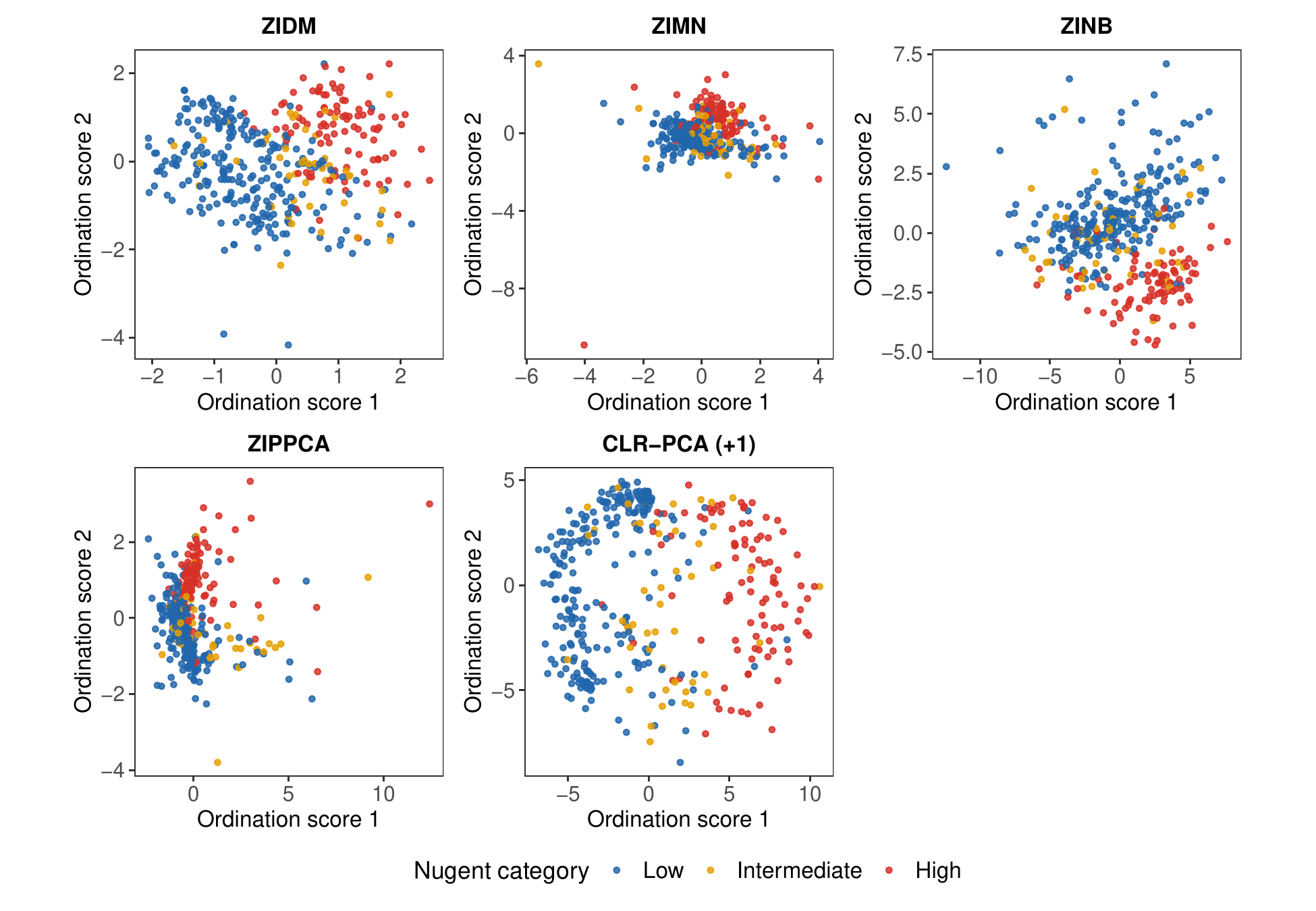}%
        }
        \caption{
            Nugent score categories: low (0--3),
            intermediate (4--6), and high (7--10).
        }
        \label{fig:nugent_category}
    \end{subfigure}

    \caption{
        Two-dimensional ordinations of the Nugent data
        obtained from ZIDM-LVM, ZIMN-LVM, ZINB-LVM, ZIPPCA, and CLR-PCA.
        The same coordinates are colored by
        (a) composition-defined community groups and
        (b) Nugent score categories.
    }
    \label{fig:nugent_ordination}
\end{figure}

\textcolor{black}{
Figures~\ref{fig:xnull_ordination} and
\ref{fig:nugent_ordination} display 
two-dimensional ordinations plot. For all ordination plots, each score matrix was independently
centered and rotated using its right singular vectors,
without unit-variance standardization or alignment across
methods.}

\textcolor{black}{
Figure~\ref{fig:xnull_ordination} shows the ordinations
of the microbial data, with samples colored by sampling
region. The four probabilistic models show some degree
of regional separation, whereas the geographic structure
is less clearly represented by CLR-PCA. Although
substantial overlap remains among the three regions,
samples from Kilpisj\"arvi tend to be more separated
from those from Mayrhofen and Ny-\AA lesund.
This pattern is consistent with the regional differences
in bacterial community composition reported by
\citet{Kumar2017}.}

The ordinations obtained from Nugent data were displayed using two sets of labels:
the five community groups in
Figure~\ref{fig:nugent_community} and the Nugent score
categories in Figure~\ref{fig:nugent_category}.
Nugent scores were categorized as low (0--3),
intermediate (4--6), and high (7--10).
Because the community groups were defined from microbial
composition, their separation reflects the representation
of composition-based structure, whereas Nugent scores
provide an external measure of vaginal microbial status.

For the Nugent data, ZIDM-LVM shows relatively clear
separation among the major community groups
(Figure~\ref{fig:nugent_community}) and a separation
trend between samples with low and high Nugent scores
(Figure~\ref{fig:nugent_category}), consistent with
the association between community composition and
Nugent scores reported by \citet{ravel2011vaginal}.

\subsection{Goodness of fit}

To assess whether the estimated compositions adequately
represented the observed microbial profiles, we multiplied
the compositions obtained from
Equation~\eqref{eq:zidm_zi_composition}
by the observed sample totals to reconstruct the counts.
The reconstructed counts were then compared with the
observed counts using the following three
goodness-of-fit statistics.

Let $y_{ij}$ denote the observed count and
$\widehat{y}_{ij}=N_i\widehat{\rho}^{0}_{ij}$
the reconstructed count, where
$N_i=\sum_{r=1}^{p}y_{ir}$ is the observed sample total
and $\widehat{\rho}^{0}_{ij}$ is the estimated composition
defined by Equation~\eqref{eq:zidm_zi_composition}.

The mean absolute range-normalized error (MARNE)
is defined as
\[
    \mathrm{MARNE}
    =
    \frac{1}{n}\sum_{i=1}^{n}
    \left(
        \frac{1}{p}\sum_{j=1}^{p}
        \frac{|y_{ij}-\widehat{y}_{ij}|}
        {\max_{1\leq r\leq p}y_{ir}
         -\min_{1\leq r\leq p}y_{ir}}
    \right).
\]
A smaller MARNE indicates closer agreement between
the observed and reconstructed counts.

The global correlation (gCOR) is the Pearson
correlation across all count entries:
\[
    \mathrm{gCOR}
    =
    \operatorname{Cor}
    \bigl(\operatorname{vec}(\widehat{\mathbf{Y}}),
          \operatorname{vec}(\mathbf{Y})\bigr).
\]

The mean taxon-specific correlation (mCOR)
is defined as
\[
    \mathrm{mCOR}
    =
    \frac{1}{p}\sum_{j=1}^{p}
    \operatorname{Cor}
    (\widehat{\boldsymbol{y}}_{\cdot j},
     \boldsymbol{y}_{\cdot j}),
\]
where $\boldsymbol{y}_{\cdot j}$ and
$\widehat{\boldsymbol{y}}_{\cdot j}$ contain the
observed and reconstructed counts for taxon $j$
across samples.
Higher gCOR and mCOR values indicate stronger
overall and within-taxon correlations, respectively.

The results are reported in Table~\ref{tab:nugent_gof}.
For the microbial data, ZIDM-LVM achieved the smallest
MARNE and the highest gCOR and mCOR.
For the Nugent data, ZIDM-LVM achieved the smallest MARNE
and the highest gCOR, indicating that its estimated
compositions yielded the closest overall count
reconstruction according to these measures.
ZINB-LVM achieved the highest mCOR for the Nugent data,
indicating the strongest average within-taxon correlation.

\begin{table}[!htbp]
    \centering
    \caption{
       Count reconstruction for the two datasets.
    Lower MARNE and higher gCOR and mCOR values indicate better fit.
    Bold values indicate the best result within each dataset and metric.
    }
    \label{tab:nugent_gof}
    \begin{tabular}{llrrr}
        \hline
        Dataset & Model & MARNE & gCOR & mCOR \\
        \hline
        Microbial & ZIDM
        & \textbf{0.008003}
        & \textbf{0.93580}
        & \textbf{0.50232} \\
        & ZIMN
        & 0.013242
        & 0.05848
        & 0.32896 \\
        & ZINB
        & 0.008531
        & 0.87020
        & 0.49347 \\
        & ZIPPCA
        & 0.008728
        & 0.92690
        & 0.47948 \\
        \hline
        Nugent & ZIDM
        & \textbf{0.01532}
        & \textbf{0.83101}
        & 0.27213 \\
        & ZIMN
        & 0.01783
        & 0.74794
        & 0.18700 \\
        & ZINB
        & 0.02057
        & 0.65769
        & \textbf{0.28411} \\
        & ZIPPCA
        & 0.01872
        & 0.72941
        & 0.25836 \\
        \hline
    \end{tabular}
\end{table}

\section{Discussion}
\label{sec:discussion}
Microbiome count data present several challenges for statistical
modeling. Due to the finite sequencing depth, the observed
counts within each sample are constrained by a fixed total,
resulting in compositional data with dependence among taxa.
Although CoDA provides a well-established framework for
analyzing such data, many traditional CoDA methods rely on
log-ratio transformations that cannot be directly applied
in the presence of zeros. This limitation is particularly
relevant for microbiome data, which are often highly sparse.
High dimensionality and overdispersion further complicate
their statistical analysis.

To account for the compositional structure of microbial counts
and the different statistical origins of observed zeros, we
incorporate zero-inflation indicators into the multinomial 
and Dirichlet–multinomial distributions and extend them within the GLLVM framework, leading to the ZIMN-LVM and ZIDM-LVM models. GLLVMs
provide a flexible framework for analyzing high-dimensional
multivariate data by using a small number of latent variables
to capture residual dependence among taxa, while allowing
environmental covariates to explain systematic variation
in microbial abundances. The proposed models therefore
combine compositional count modeling, explicit accommodation
of excess zeros, and a low-dimensional representation of
microbial communities, while allowing the effects of observed
covariates to be estimated. 
For likelihood-based estimation and model-based ordination,
we developed two variational approximation approaches:
a conventional VA that retains uncertainty in the
structural-zero indicators and a classification VA
that uses binary assignments for these indicators.

We conducted simulation studies to evaluate composition
recovery, parameter estimation, and ordination recovery,
comparing the proposed models with commonly used approaches
for multivariate count data. Overall, ZIDM-LVM showed competitive
performance in composition recovery and parameter estimation
across different generating models, particularly in estimating
regression coefficients and zero-inflation probabilities.
However, its ordination recovery was less consistent across
different generating models. 
\textcolor{black}{The results also highlighted the importance of
accounting for the distinct statistical meanings of
sampling and structural zeros, rather than simply
replacing zeros with small positive values,
particularly in highly sparse settings.
}

We further applied the proposed models to two real microbiome
datasets to investigate their practical applicability. The
results demonstrated their ability to represent the observed
count structure and explore underlying patterns in microbial
community composition.

It is worth noting that variational approximation involves
optimizing a generally non-convex objective function, making
estimation potentially sensitive to the choice of initial values
and convergence to local optima. Under extremely sparse settings,
estimation may become less reliable.
Furthermore, the subnominal confidence interval
coverage observed in the simulations indicates that uncertainty
quantification requires further investigation.
\textcolor{black}{
Similar undercoverage has been reported in previous simulation
studies of GLLVMs \citep{Nikuetal:2017,Zeng2023}. The observed undercoverage may partly be related to
variational approximation. In particular,
Hessian-based covariance estimates derived from
a variational objective function do not necessarily
capture the sampling variability of variational
estimators, and bias in variational parameter estimates
may also affect confidence interval coverage
\citep{WestlingMcCormick2019}. Numerical difficulties were also encountered in Hessian-based
standard error estimation in some simulation replicates,
resulting in unavailable interval estimates.
These difficulties may affect the reliability of
uncertainty quantification, although they do not necessarily
explain the observed undercoverage.
} \textcolor{black}{
Moreover, in constructing the ELBO, we adopted a
structured mean-field factorization that neglects
some dependencies between latent and augmentation
variables, which may also contribute to
underestimated uncertainty.
More flexible variational distributions that retain
these dependencies could potentially improve
uncertainty quantification.} Improving the stability of optimization and the reliability of statistical inference therefore remains important.

Future research could investigate more effective initialization
strategies and alternative variational approximations to improve
estimation accuracy and computational efficiency.

Alternative covariance estimation methods, such as sandwich
estimators or bootstrap approaches, could also be explored
to improve the reliability of interval estimation.

\textcolor{black}{
Linear response variational Bayes (LRVB), which provides
a correction to covariance estimates obtained from
mean-field variational approximations
\citep{Giordano2018}, may offer
another direction for future investigation.
}
Extensions to other compositional count distributions may also
provide greater flexibility in modeling overdispersion and
dependence among taxa.

\section*{Acknowledgments}
This work was supported by the Finnish Doctoral Program Network in Artificial Intelligence (AI-DOC; decision no. VN/3137/2024-OKM-6). We also acknowledge the support from the Research Council of Finland (356484) to ST, the support from the Research Council of Finland (363261) to KN, and CSC – IT Center for Science, Finland, for providing computational resources.

\appendix
\section{Zero-inflated multinomial latent variable model}
\label{app:zimn}
\subsection{Model formulation and Gamma augmentation}
\label{app:zimn_model}

We first formulate the multinomial component of the proposed
ZIMN-LVM, temporarily ignoring zero inflation. Let
$\mathbf Y_i=(y_{i1},\ldots,y_{ip})^\top$ denote the vector of taxon
counts for sample $i$, with $\sum_{j=1}^p y_{ij}=N_i$. We assume
\begin{equation*}
\label{eq:zimn_multinomial}
\mathbf Y_i\mid N_i,\boldsymbol{\rho}_i
\sim
\operatorname{Multinomial}(N_i,\boldsymbol{\rho}_i).
\end{equation*}

The multinomial GLLVM is specified as
\begin{equation}
\label{eq:zimn_gllvm}
\begin{aligned}
p(\mathbf Y_i;N_i,\boldsymbol{\rho}_i)
&=
\binom{N_i}{y_{i1},\ldots,y_{im}}
\prod_{j=1}^p\rho_{ij}^{y_{ij}},
\\
\rho_{ij}
&=
\frac{\lambda_{ij}}
{\sum_{k=1}^p\lambda_{ik}},
\qquad j=1,\ldots,p,
\\
\lambda_{ij}
&=
\exp(\eta_{ij}),
\qquad
\eta_{ij}
=
\alpha_i+\beta_{0j}
+\mathbf x_i^\top\boldsymbol{\beta}_j
+\mathbf u_i^\top\boldsymbol{\gamma}_j,
\\
\mathbf u_i
&\sim
N_d(\mathbf 0,\mathbf I_d).
\end{aligned}
\end{equation}
Here, $\boldsymbol{\rho}_i=(\rho_{i1},\ldots,\rho_{im})^\top$
denotes the underlying composition, satisfying
$\sum_{j=1}^p\rho_{ij}=1$, while $\lambda_{ij}>0$ denotes the
corresponding positive intensity.
Substituting the compositional probabilities in
\eqref{eq:zimn_gllvm} into the multinomial likelihood gives
\begin{align*}
p(\mathbf y_i\mid\mathbf u_i)
&=
\binom{N_i}{y_{i1},\ldots,y_{im}}
\prod_{j=1}^p
\left(
\frac{\lambda_{ij}}
{\sum_{k=1}^p\lambda_{ik}}
\right)^{y_{ij}}
\nonumber\\
&=
\binom{N_i}{y_{i1},\ldots,y_{im}}
\left(
\prod_{j=1}^p\lambda_{ij}^{y_{ij}}
\right)
\left(
\sum_{j=1}^p\lambda_{ij}
\right)^{-N_i},
\label{eq:zimn_mn_likelihood}
\end{align*}

The normalization term
$\left(\sum_{j=1}^p\lambda_{ij}\right)^{-N_i}$
couples the taxon-specific intensities within each sample. Following
\citet{menezes2025:zanim}, we introduce the latent variable
$\phi_i\mid\mathbf u_i
\sim
\operatorname{Gamma}
\left(N_i,\sum_{j=1}^p\lambda_{ij}\right)$,
where the second parameter denotes the rate. Using the Gamma integral
identity, the coupled normalization term can be written as
\begin{equation*}
\label{eq:zimn_gamma_representation}
\left(
\sum_{j=1}^p\lambda_{ij}
\right)^{-N_i}
=
\frac{1}{\Gamma(N_i)}
\int_0^\infty
\phi_i^{N_i-1}
\exp\left\{
-\phi_i\sum_{j=1}^p\lambda_{ij}
\right\}
\,d\phi_i.
\end{equation*}
This gives the augmented joint likelihood
\begin{equation}
\label{eq:aug_mn1}
p(\mathbf Y_i,\phi_i;\boldsymbol{\lambda}_i,N_i)
=
\frac{N_i!\phi_i^{N_i-1}}{\Gamma(N_i)}
\prod_{j=1}^p
\frac{
\lambda_{ij}^{y_{ij}}
e^{-\lambda_{ij}\phi_i}
}{
y_{ij}!
}.
\end{equation}
Integrating $\phi_i$ out of \eqref{eq:aug_mn1} recovers the original
multinomial likelihood. Importantly, the augmentation allows the
individual taxon-specific likelihood contributions to take a
Poisson-like form \citep{menezes2025:zanim},
\begin{equation*}
\label{eq:aug_mn2}
p(y_{ij},\phi_i;\lambda_{ij})
\propto
\frac{
\lambda_{ij}^{y_{ij}}
e^{-\lambda_{ij}\phi_i}
}{
y_{ij}!
},
\qquad
y_{ij}\in\{0,\ldots,N_i\}.
\end{equation*}

We now introduce zero inflation for observed zeros through a latent
indicator $z_{ij}$, where $z_{ij}=1$ denotes the structural-zero
component and $z_{ij}=0$ denotes the multinomial sampling component,
\begin{equation*}
\label{eq:zimn_indicator}
z_{ij}
\sim
\operatorname{Bernoulli}(\zeta_j),
\qquad
P(z_{ij}=1)=\zeta_j,
\end{equation*}
with $0<\zeta_j<1$. When $y_{ij}=0$, the observation may therefore
arise either from the structural-zero component ($z_{ij}=1$) or as a
sampling zero from the multinomial component ($z_{ij}=0$).  

When $z_{ij}=0$, taxon $j$ belongs to the multinomial sampling
component with intensity $\lambda_{ij}=\exp(\eta_{ij})$, whereas
$z_{ij}=1$ sets the corresponding intensity to zero. Conditional on
$\mathbf z_i=(z_{i1},\ldots,z_{im})^\top$, the resulting compositional
probabilities are therefore
\begin{equation*}
\label{eq:zimn_zi_composition}
\rho_{ij}(\mathbf z_i)
=
\frac{
(1-z_{ij})\lambda_{ij}
}{
\sum_{k=1}^p (1-z_{ik})\lambda_{ik}
}
=
\frac{
(1-z_{ij})\exp(\eta_{ij})
}{
\sum_{k=1}^p (1-z_{ik})\exp(\eta_{ik})
}.
\end{equation*}
It can be seen that when a component is a structural zero,
$z_{ij}=1$ removes the corresponding taxon-specific intensity and
forces $\rho_{ij}(\mathbf z_i)=0$, whereas the remaining positive
intensities are renormalized to sum to one.

Conditional on $z_{ij}$, the taxon-wise augmented contribution can be
written as
\begin{equation}
\label{eq:zimn}
p(y_{ij},\phi_i;\lambda_{ij}\mid z_{ij})
\propto
\begin{cases}
\displaystyle
\frac{
\lambda_{ij}^{y_{ij}}
e^{-\lambda_{ij}\phi_i}
}{
y_{ij}!
},
& z_{ij}=0,\\[4mm]
\delta_0(y_{ij}),
& z_{ij}=1,
\end{cases}
\end{equation}
where $\delta_0(y_{ij})$ denotes a point mass at zero, assigning
probability one to $y_{ij}=0$ and zero probability to any positive
count under the structural-zero component.

The two components in \eqref{eq:zimn} can be represented
jointly as
\begin{equation*}
\label{eq:zimn_complete_factor}
\left[
(1-\zeta_j)
\frac{
\exp\{y_{ij}\eta_{ij}-\phi_i\exp(\eta_{ij})\}
}{
y_{ij}!
}
\right]^{1-z_{ij}}
\left[
\zeta_j\delta_0(y_{ij})
\right]^{z_{ij}}.
\end{equation*}
Accordingly, the complete-data
log-likelihood is
\begin{align*}
\label{eq:compll}
\log \mathcal L
(\boldsymbol{\Psi};
 \mathbf y,\boldsymbol{\phi},\mathbf z,\mathbf u)
={}&
\sum_{i=1}^n
\Bigg\{
\log(N_i!)
+
(N_i-1)\log\phi_i
-
\log\Gamma(N_i)
+
\log p(\mathbf u_i)
\nonumber\\
&\qquad
+
\sum_{j=1}^p
(1-z_{ij})
\left[
\log(1-\zeta_j)
+
y_{ij}\eta_{ij}
-
\phi_i\exp(\eta_{ij})
-
\log(y_{ij}!)
\right]
\nonumber\\
&\qquad
+
\sum_{j:y_{ij}=0}
z_{ij}\log\zeta_j
\Bigg\}.
\end{align*}

\subsection{Variational approximation}

To construct the variational lower bound for the ZIMN-LVM in a form
analogous to \eqref{eq:general_elbo}, we introduce variational
distributions for the latent and augmented quantities. The latent and
augmented quantities in the ZIMN-LVM are the latent variables $\mathbf u$,
the Gamma augmentation variables $\boldsymbol{\phi}$, and the
zero-inflation indicators $\mathbf z$. 
The variational lower bound for the ZIMN-LVM is as follow

\begin{equation}
\label{eq:zimn_elbo_expectation}
\begin{aligned}
\operatorname{ELBO}_{\mathrm{ZIMN},i}
(\boldsymbol{\Psi},\boldsymbol{\xi}\mid q)
={}&
\mathbb E_q
\left[
\log
p(
\mathbf y_i,
\mathbf u_i,
\phi_i,
\mathbf z_i
\mid\boldsymbol{\Psi})
\right]\\
&-
\mathbb E_q
\left[
\log
q(
\mathbf u_i,
\phi_i,
\mathbf z_i;
\boldsymbol{\xi})
\right].
\end{aligned}
\end{equation}

Under the mean-field assumption
introduced in the general variational framework, we take
\begin{equation*}
\label{eq:zimn_meanfield}
q(\mathbf u,\boldsymbol{\phi},\mathbf z)
=
q(\mathbf u)q(\boldsymbol{\phi})q(\mathbf z)
=
\prod_{i=1}^n
q(\mathbf u_i)q(\phi_i)
\prod_{j=1}^p q(z_{ij}),
\end{equation*}
where
\begin{equation}
\label{eq:zimn_variational_distributions}
\begin{gathered}
q(\mathbf u_i) \sim N_d(\mathbf a_i,\mathbf A_i),
\qquad
q(\phi_i) \sim \operatorname{Gamma}(N_i,S_i),\\[4pt]
q(z_{ij}) \sim \operatorname{Bernoulli}(\widetilde{\pi}_{ij}),
\qquad y_{ij}=0.
\end{gathered}
\end{equation}
We define
$\boldsymbol{\xi}
=
\{\mathbf a_i,\mathbf A_i,S_i,\widetilde{\pi}_{ij}\}$
as variational parameters, and
$\boldsymbol{\Psi}
=
\{\alpha_i,\beta_{0j},\boldsymbol{\beta}_j,
\boldsymbol{\gamma}_j,\zeta_j\}$
as model parameters.

From \eqref{eq:zimn_variational_distributions},
$\mathbb E_q(\phi_i)=N_i/S_i$. For the exponential term, we include
the second-order correction induced by the variational covariance
$\mathbf A_i$, giving
\begin{equation*}
\label{eq:zimn_exp_eta}
\mathbb E_q\left[\exp(\eta_{ij})\right]
=
\exp\left\{
\widetilde{\eta}_{ij}
+
\frac{1}{2}
\boldsymbol{\gamma}_j^\top
\mathbf A_i
\boldsymbol{\gamma}_j
\right\},
\end{equation*}
where
$\widetilde{\eta}_{ij}
=
\alpha_i+\beta_{0j}
+\mathbf x_i^\top\boldsymbol{\beta}_j
+\mathbf a_i^\top\boldsymbol{\gamma}_j$
is the linear predictor evaluated at the variational mean
$\mathbf a_i$.
Given the variational distributions of $\mathbf u_i$ and $\phi_i$
specified above, we can further obtain the variational classification
probability $\widetilde{\pi}_{ij}$ for the zero-inflation indicator.
Since a positive count necessarily belongs to the multinomial sampling
component, while an observed zero may arise from either component, we
have
\begin{equation*}
\label{eq:zimn_pi}
\widetilde{\pi}_{ij}
=
\begin{cases}
0,
& y_{ij}>0,\\[8pt]
\displaystyle
\frac{
\zeta_j
}{
\zeta_j
+
(1-\zeta_j)\exp(-\widetilde M_{ij})
},
& y_{ij}=0,
\end{cases}
\end{equation*}
where
$\widetilde M_{ij}
=
(N_i/S_i)
\exp\{
\widetilde{\eta}_{ij}
+
\tfrac{1}{2}
\boldsymbol{\gamma}_j^\top
\mathbf A_i
\boldsymbol{\gamma}_j
\}$.
Here, $\zeta_j$ denotes the model-level structural-zero probability,
whereas $\widetilde{\pi}_{ij}$ denotes the variational probability
of membership in the structural-zero component.

Substituting these variational quantities into
\eqref{eq:zimn_elbo_expectation} and collecting the corresponding
entropy terms yields the ZIMN-LVM variational lower bound. Up to an
additive constant independent of the model and variational
parameters,
\begin{align*}
\label{eq:zimn_elbo}
\operatorname{ELBO}_{\mathrm{ZIMN}}
={}&
\frac{1}{2}
\sum_{i=1}^n
\left\{
\log|\mathbf A_i|
-
\operatorname{tr}(\mathbf A_i)
-
\mathbf a_i^\top\mathbf a_i
-
2N_i\log S_i
\right\}
\nonumber\\
&+
\sum_{i=1}^n
\sum_{j=1}^p
(1-\widetilde{\pi}_{ij})
\left[
\log(1-\zeta_j)
+
y_{ij}\widetilde{\eta}_{ij}
-
\widetilde M_{ij}
-
\log(1-\widetilde{\pi}_{ij})
\right]
\nonumber\\
&+
\sum_{i=1}^n
\sum_{j:y_{ij}=0}
\widetilde{\pi}_{ij}
\left[
\log\zeta_j
-
\log\widetilde{\pi}_{ij}
\right].
\end{align*}

Following
\eqref{eq:general_va_estimation}, the model and variational parameters
are estimated jointly by maximizing
$\mathcal L_{\mathrm{ZIMN}}$ with respect to
$\boldsymbol{\Psi}$ and $\boldsymbol{\xi}$.

\section{\textcolor{black}{Simulation under the ZIMN-LVM data-generating mechanism}}
\label{app:zimn_dgm}

For the ZIMN-LVM, the latent variable $\mathbf{u}_i$ is independently
generated from $N_d(\mathbf{0},\mathbf{I}_d)$ for each sample $i$,
while the structural-zero indicator $z_{ij}$ is independently
generated from $\mathrm{Bernoulli}(\zeta_j)$, where $z_{ij}=1$
indicates a structural zero.

The linear predictor is specified as
\[
\eta_{ij}
=
\beta_{0j}
+\mathbf{x}_i^\top\boldsymbol{\beta}_j
+\mathbf{u}_i^\top\boldsymbol{\gamma}_j.
\]
The underlying composition is obtained by normalizing the
exponentiated linear predictors as
$\rho_{ij}
=
\frac{\exp(\eta_{ij})}
{\sum_{\ell=1}^{p}\exp(\eta_{i\ell})}$.

The observed count vector $\mathbf{Y}_i$ is generated according to
\[
\mathbf{Y}_i
\sim
\begin{cases}
\delta_{\mathbf{0}_p}(\cdot),
& z_{ij}=1,\ \forall j,\\[1mm]
\mathrm{Multinomial}(N_i,\rho_{i1}^{0},\ldots,\rho_{ip}^{0}),
& \text{otherwise},
\end{cases}
\]
where
$\rho_{ij}^{0}
=
\frac{(1-z_{ij})\rho_{ij}}
{\sum_{\ell=1}^{p}(1-z_{i\ell})\rho_{i\ell}}$.

The common parameter settings for the ZIMN-LVM data-generating mechanism
are the same as those reported in Table~\ref{tab:simulation_parameters}. The ZIMN-LVM specific zero-inflation and sampling-intensity
settings are shown in Table~\ref{tab:simulation_sparsity_zimn}. \textcolor{black}{No additional sample-specific intercept $\alpha_i$ was included
in the ZIMN-LVM data-generating mechanism, since a common
additive intercept cancels out under the softmax normalization.}

\begin{table}[htbp]
\centering
\caption{Zero-inflation and calibrated sampling-intensity settings for ZIMN-LVM.}
\label{tab:simulation_sparsity_zimn}
\small
\renewcommand{\arraystretch}{1.20}
\setlength{\tabcolsep}{8pt}

\begin{tabular}{lc|c|c}
\toprule
\multicolumn{2}{c|}{\textbf{Zero inflation}}
&
\multicolumn{1}{c|}{$(n,p)=(50,100)$}
&
\multicolumn{1}{c}{$(n,p)=(100,50)$}
\\
\cmidrule(lr){1-2}
\cmidrule(lr){3-3}
\cmidrule(lr){4-4}

\textbf{ZI level}
&
\textbf{$\zeta_j$}
&
\textbf{ZIMN $N_i$}
&
\textbf{ZIMN $N_i$}
\\
\midrule

Low
&
$\mathcal{U}(0.05,0.15)$
&
$\mathcal{U}_d(200,300)$
&
$\mathcal{U}_d(100,150)$
\\

Medium
&
$\mathcal{U}(0.20,0.30)$
&
$\mathcal{U}_d(320,480)$
&
$\mathcal{U}_d(140,210)$
\\

High
&
$\mathcal{U}(0.35,0.45)$
&
$\mathcal{U}_d(800,1200)$
&
$\mathcal{U}_d(400,600)$
\\

\bottomrule
\end{tabular}

\vspace{2mm}

\begin{minipage}{0.90\linewidth}

\end{minipage}

\end{table}

\begin{figure}[!htbp]
    \centering
    \makebox[\textwidth][c]{%
        \includegraphics[
            width=1\textwidth,
            keepaspectratio
        ]{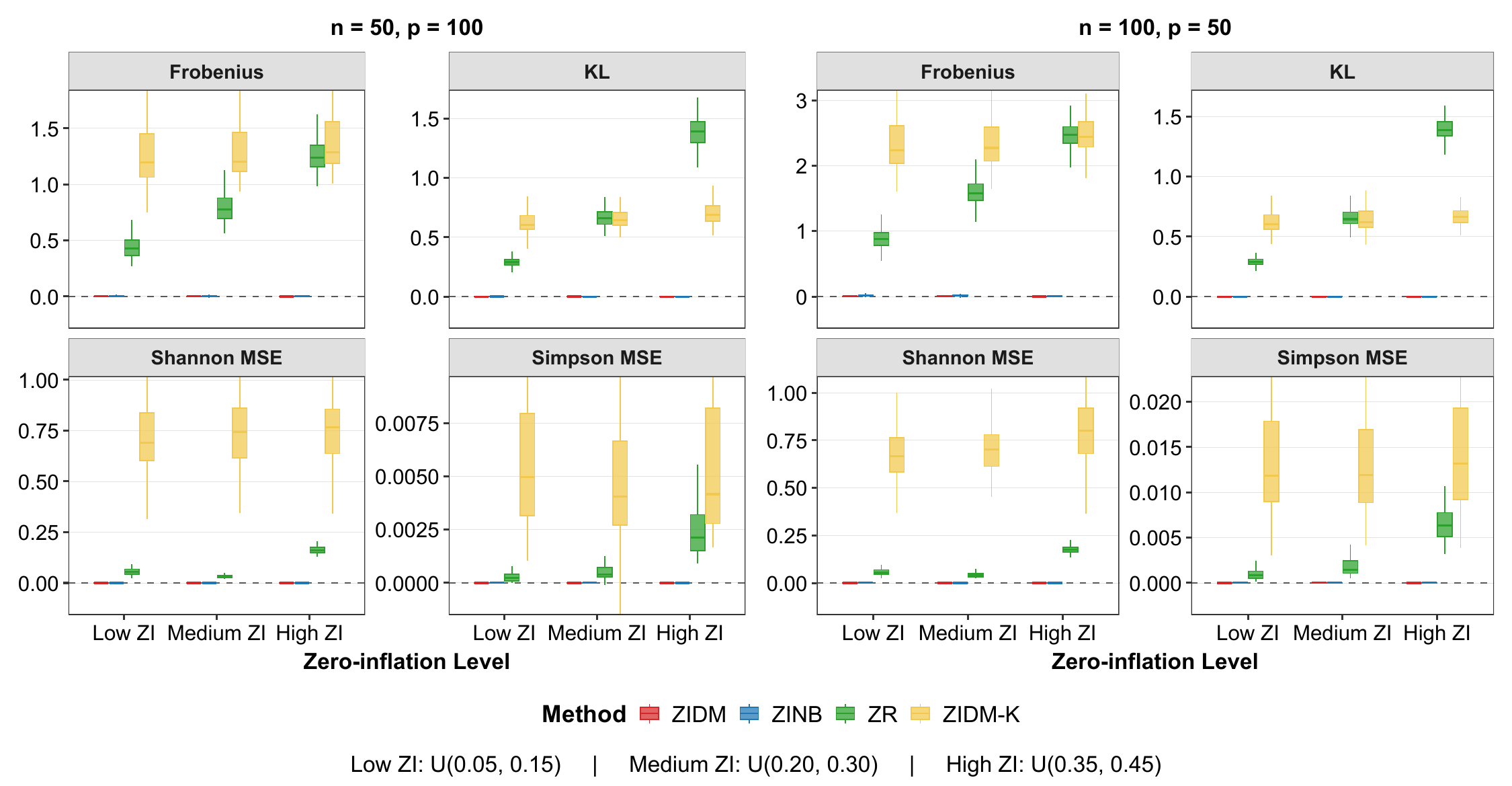}
    }
    \caption{Differences in composition recovery errors relative
    to the generating model for ZIMN-LVM data across three levels
    of structural zero inflation and two dimensional settings.
    Positive values indicate larger recovery errors.}
    \label{fig:zimn_composition}
\end{figure}

From  the Figure~\ref{fig:zimn_composition} three LVM-based methods show comparable and consistently strong performance in
recovering the underlying composition, outperforming ZIDM-K and ZR. As the proportion of structural zeros increased, ZR was the most affected method.
Table~\ref{tab:ordination_recovery_zimn} shows comparable
ordination recovery performance across the three methods,
with ZIMN-LVM achieving slightly lower Procrustes errors overall.

\begin{table}[!htbp]
\centering
\caption{Mean Procrustes error for ordination recovery under the ZIMN-LVM
data-generating mechanism. Smaller values indicate better recovery
of the true latent structure.}
\label{tab:ordination_recovery_zimn}

\small
\setlength{\tabcolsep}{5.0pt}
\renewcommand{\arraystretch}{1.12}

\begin{tabular}{lccc@{\hspace{8mm}}ccc}
\toprule
& \multicolumn{3}{c}{$(n,p)=(50,100)$}
& \multicolumn{3}{c}{$(n,p)=(100,50)$} \\
\cmidrule(lr){2-4}
\cmidrule(lr){5-7}

& ZIDM & ZIMN & ZINB
& ZIDM & ZIMN & ZINB \\
\midrule

\multicolumn{7}{c}{ZIMN-LVM generator} \\
\midrule

Low
& 5.74 & 5.64 & \textbf{5.59}
& 8.53 & \textbf{8.30} & 8.50 \\

Medium
& 4.88 & \textbf{4.80} & 4.81
& 7.58 & \textbf{7.05} & 7.22 \\

High
& 5.20 & \textbf{5.15} & \textbf{5.15}
& 7.06 & \textbf{6.10} & 6.15 \\

\bottomrule
\end{tabular}
\end{table}

\begin{figure}[!htbp]
    \centering
    \includegraphics[
        width=1\textwidth
    ]{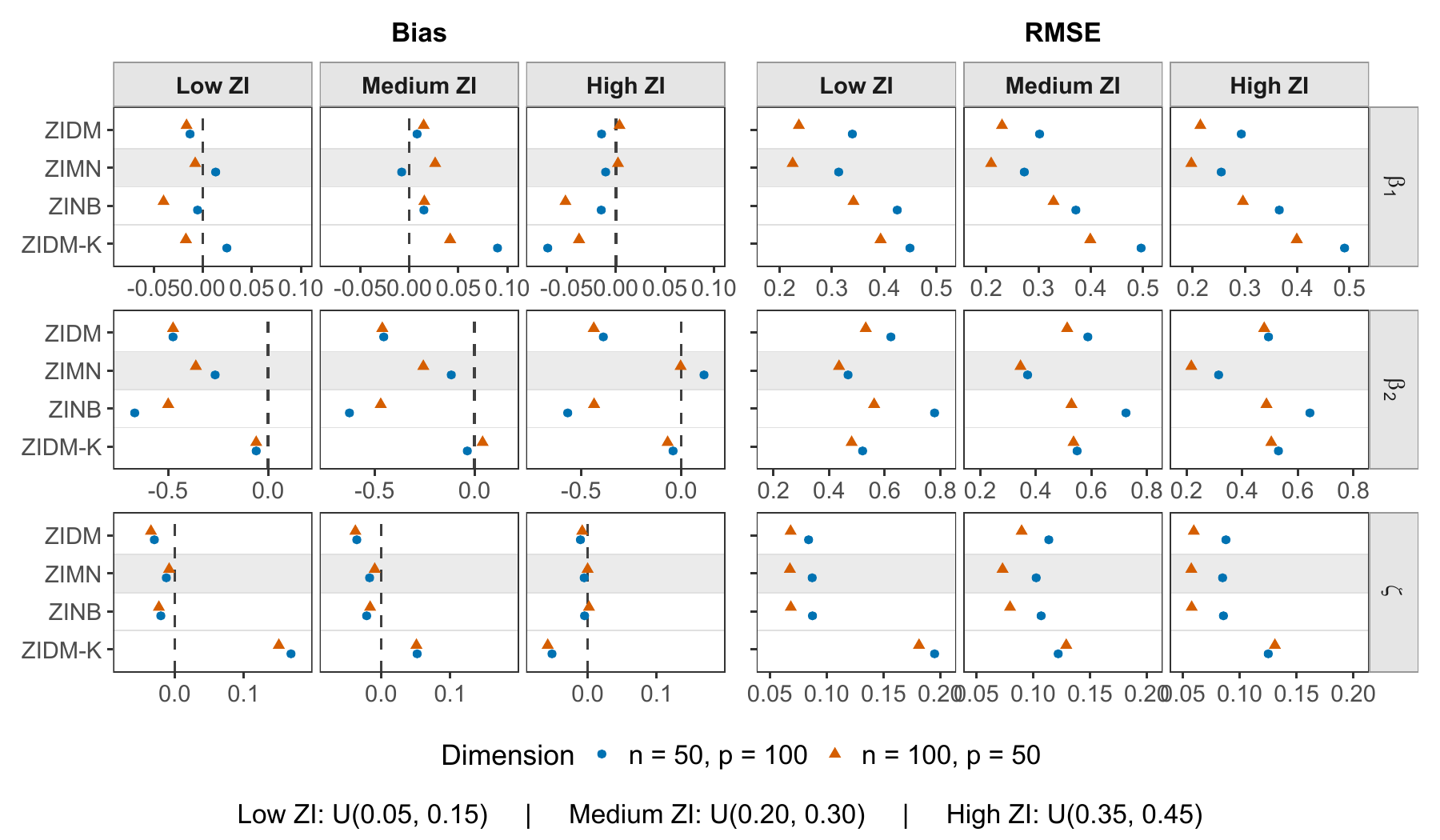}

    \caption{Mean Bias and RMSE of parameter estimates for data
    generated from the ZIMN-LVM across three levels of structural
    zero inflation and two dimensional settings.
    Gray shading highlights the method corresponding to the
    generating model. The dashed line indicates zero bias.}
    \label{fig:parameter_zimn}
\end{figure}
Figure~\ref{fig:parameter_zimn} and Table~\ref{tab:ci_recovery_zimn}
show that ZIMN-LVM generally achieved the best parameter estimation
performance, followed closely by ZIDM-LVM, with relatively small
biases and RMSEs and good confidence interval coverage.

\begin{table}[!htbp]
\centering
\caption{Mean confidence interval (CI) width and empirical
coverage probability for the regression coefficient estimates
for data generated from ZIMN-LVM. Values are reported
as CI width / \textbf{coverage}.}
\label{tab:ci_recovery_zimn}

\scriptsize
\renewcommand{\arraystretch}{1.15}
\setlength{\tabcolsep}{1.5pt}

\begin{tabular}{@{}l@{\hspace{3pt}}l
                *{4}{c}@{\hspace{3mm}}*{4}{c}@{}}
\toprule
& &
\multicolumn{4}{c}{$(n,p)=(50,100)$} &
\multicolumn{4}{c}{$(n,p)=(100,50)$} \\
\cmidrule(lr){3-6}
\cmidrule(lr){7-10}

ZI & Param.
& ZIDM & ZIMN & ZINB & ZIDM-K
& ZIDM & ZIMN & ZINB & ZIDM-K \\
\midrule

\multicolumn{10}{c}{ZIMN-LVM generator} \\
\midrule

Low
& $\beta_1$
& 1.28/\textbf{0.94}
& 1.13/\textbf{0.92}
& 1.22/\textbf{0.82}
& 0.85/\textbf{0.66}
& 0.87/\textbf{0.92}
& 0.76/\textbf{0.90}
& 0.89/\textbf{0.78}
& 0.58/\textbf{0.58} \\

& $\beta_2$
& 1.53/\textbf{0.73}
& 1.36/\textbf{0.85}
& 1.41/\textbf{0.52}
& 0.99/\textbf{0.59}
& 1.00/\textbf{0.52}
& 0.89/\textbf{0.59}
& 0.95/\textbf{0.45}
& 0.67/\textbf{0.46} \\

\addlinespace[2pt]

Medium
& $\beta_1$
& 1.23/\textbf{0.95}
& 1.06/\textbf{0.94}
& 1.15/\textbf{0.85}
& 0.83/\textbf{0.63}
& 0.86/\textbf{0.93}
& 0.74/\textbf{0.91}
& 0.87/\textbf{0.79}
& 0.59/\textbf{0.56} \\

& $\beta_2$
& 1.44/\textbf{0.72}
& 1.26/\textbf{0.92}
& 1.31/\textbf{0.51}
& 0.97/\textbf{0.58}
& 0.95/\textbf{0.49}
& 0.82/\textbf{0.70}
& 0.88/\textbf{0.45}
& 0.65/\textbf{0.41} \\

\addlinespace[2pt]

High
& $\beta_1$
& 1.14/\textbf{0.93}
& 0.93/\textbf{0.91}
& 1.03/\textbf{0.81}
& 0.86/\textbf{0.65}
& 0.78/\textbf{0.91}
& 0.64/\textbf{0.85}
& 0.73/\textbf{0.73}
& 0.61/\textbf{0.58} \\

& $\beta_2$
& 1.26/\textbf{0.73}
& 1.05/\textbf{0.87}
& 1.09/\textbf{0.47}
& 0.98/\textbf{0.61}
& 0.85/\textbf{0.46}
& 0.70/\textbf{0.85}
& 0.74/\textbf{0.41}
& 0.67/\textbf{0.47} \\

\bottomrule
\end{tabular}
\end{table}

\section{Additional simulation results}
\label{app:additional_simulation}

\subsection{Conventional VA versus classification VA}
\label{app:soft_hard}

\begin{figure}[!htbp]
    \centering

    \makebox[\textwidth][c]{%
        \includegraphics[
            width=1.08\textwidth
        ]{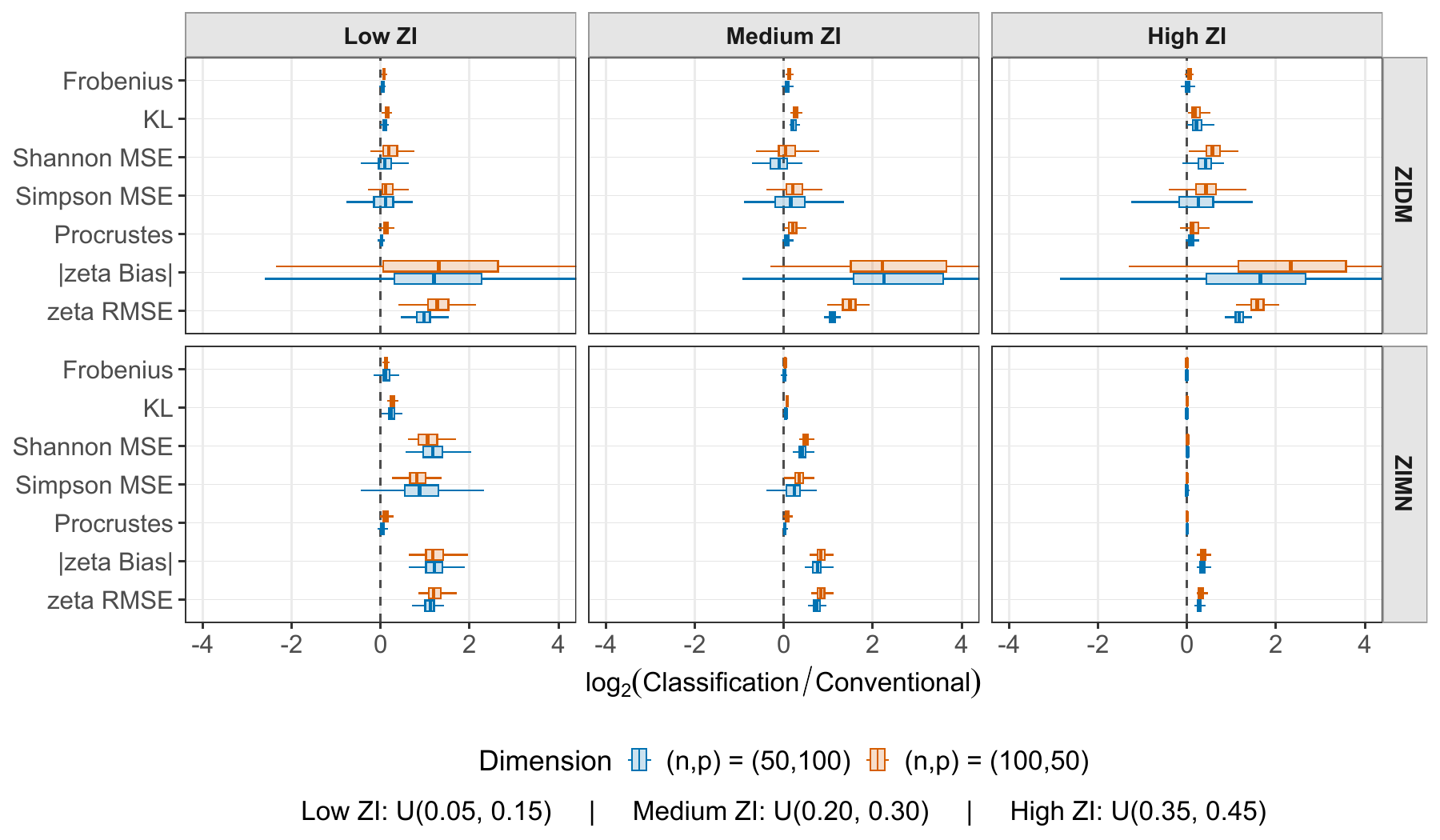}
    }

    \vspace{2mm}
    {\small (a) ZIDM-LVM generator}

    \vspace{4mm}

    \makebox[\textwidth][c]{%
        \includegraphics[
            width=1.08\textwidth
        ]{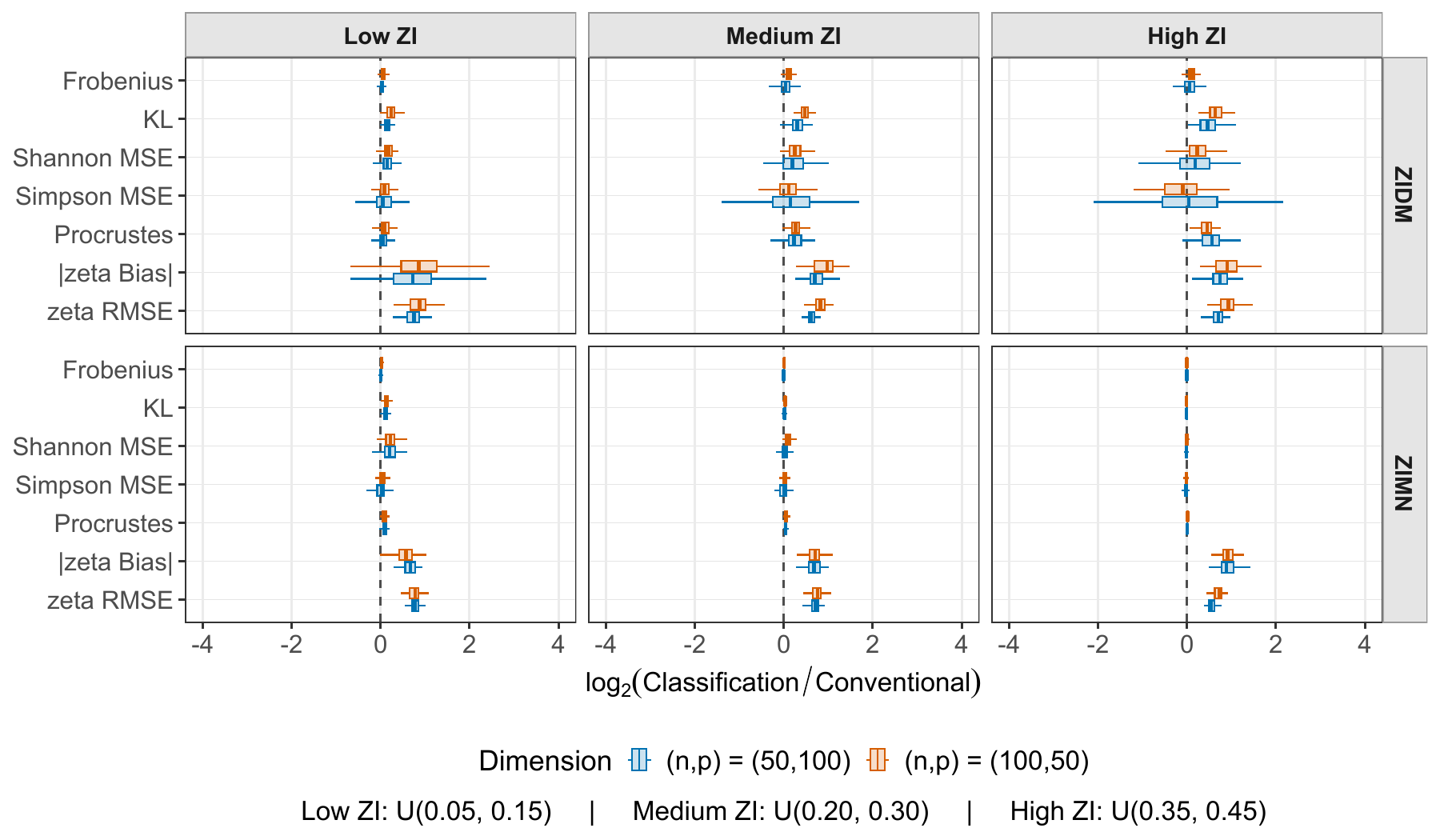}
    }

    \vspace{2mm}
    {\small (b) ZINB-LVM generator}

    \caption{Log-ratios of recovery errors between classification VA and
conventional VA for data generated from (a) ZIDM-LVM and
(b) ZINB-LVM. Values are calculated as
$\log_2(\mathrm{Classification\ VA}/\mathrm{Conventional\ VA})$,
with positive values favoring conventional VA.
}
    \label{fig:appendix_soft_hard_relative}
\end{figure}

In addition to conventional VA, which was used in the main
analysis, we also considered classification VA for both
ZIDM-LVM and ZIMN-LVM. Classification VA uses hard assignments
for the structural-zero indicators, whereas conventional VA
retains their variational probabilities throughout optimization.
Figure~\ref{fig:appendix_soft_hard_relative}
compares the recovery errors of the two approaches, with bias
and RMSE evaluated for the zero-inflation parameter $\zeta$.
Overall, conventional VA generally provided better recovery
than classification VA, with the largest differences observed
in the absolute bias of $\zeta$. In contrast, the differences
in composition and ordination recovery were relatively small.

Table~\ref{tab:appendix_soft_hard_runtime} reports the
differences in mean computation time between conventional VA
and classification VA. Classification VA was generally faster,
whereas conventional VA provided more accurate estimates
of $\zeta$, supporting its use in the main analysis.

\begin{table}[!htbp]
\centering
\caption{Differences in mean computation time (seconds)
between conventional VA and classification VA for
ZIDM-LVM and ZIMN-LVM. Positive values indicate longer
computation times for conventional VA.}
\label{tab:appendix_soft_hard_runtime}

\small
\setlength{\tabcolsep}{8pt}
\renewcommand{\arraystretch}{1.12}

\begin{tabular}{lcc@{\hspace{8mm}}cc}
\toprule
& \multicolumn{2}{c}{$(n,p)=(50,100)$}
& \multicolumn{2}{c}{$(n,p)=(100,50)$} \\
\cmidrule(lr){2-3}
\cmidrule(lr){4-5}

ZI level
& ZIDM & ZIMN
& ZIDM & ZIMN \\
\midrule

\multicolumn{5}{c}{ZIDM-LVM generator} \\
\midrule

Low
& 3.60 & 1.19
& 1.36 & 1.47 \\

Medium
& 2.43 & 0.76
& 1.13 & 5.43 \\

High
& 1.30 & -1.46
& 0.27 & 0.85 \\

\midrule

\multicolumn{5}{c}{ZINB-LVM generator} \\
\midrule

Low
& 12.62 & 3.28
& 2.31 & 1.12 \\

Medium
& 4.85 & 2.10
& 2.22 & 0.93 \\

High
& 5.62 & 1.67
& 5.19 & 1.24 \\

\bottomrule
\end{tabular}
\end{table}

\FloatBarrier

\section{Notation and abbreviations}
\label{app:notation}

This appendix summarizes the main mathematical notation
and abbreviations used throughout the paper.
In figures and tables, ZIMN and ZIDM are used as
short labels for ZIMN-LVM and ZIDM-LVM, respectively.
\subsection{Notation}

\small
\renewcommand{\arraystretch}{1.15}
\setlength{\LTleft}{0pt}
\setlength{\LTright}{0pt}

\begin{longtable}{p{0.29\textwidth}p{0.65\textwidth}}

\caption{Summary of mathematical notation.}
\label{tab:notation}\\

\toprule
\textbf{Symbol} & \textbf{Description} \\
\midrule
\endfirsthead

\multicolumn{2}{l}{\textit{Table \thetable\ (continued)}}\\
\toprule
\textbf{Symbol} & \textbf{Description} \\
\midrule
\endhead

\bottomrule
\endfoot

\multicolumn{2}{l}{observed data}\\
\addlinespace[3pt]

$n$ & Number of samples \\

$p$ & Number of taxa \\

$d$ & Dimension of the latent space \\

$y_{ij}$ & Observed count of taxon $j$ in sample $i$ \\

$\mathbf Y_i$ & Count vector for sample $i$ \\

$N_i$ & Total number of reads in sample $i$ \\

$\mathbf x_i$ & Vector of observed covariates for sample $i$ \\

$\mathbf I_d$ & $d\times d$ identity matrix \\

\addlinespace[6pt]
\multicolumn{2}{l}{Model parameters and latent variables}\\
\addlinespace[3pt]

$\alpha_i$ & Sample-specific effect \\

$\beta_{0j}$ & Taxon-specific intercept \\

$\boldsymbol{\beta}_j$ &
Vector of regression coefficients for taxon $j$ \\

$\mathbf u_i$ &
$d$-dimensional latent variable vector for sample $i$ \\

$\boldsymbol{\gamma}_j$ &
Vector of latent variable loadings for taxon $j$ \\

$\eta_{ij}$ & Linear predictor \\

$\mu_{ij}$ &
Conditional mean parameter in the general GLLVM and
ZINB model \\

$\rho_{ij}$ &
Underlying composition of taxon $j$
in sample $i$ \\

$\rho_{ij}^{0}$ &
Composition after applying
structural-zero indicators \\

$\theta_{ij}$ &
Dirichlet concentration parameter in the ZIDM-LVM \\

$\lambda_{ij}$ &
Gamma latent variable in the ZIDM-LVM \\

$z_{ij}$ &
Structural-zero indicator, with $z_{ij}=1$
indicating a structural zero \\

$\zeta_j$ &
Taxon-specific structural-zero probability \\

$\kappa$ &
Negative binomial dispersion parameter in
the ZINB model \\

$\phi_i$ &
Sample-specific Gamma augmentation variable \\

$\omega_{ij}$ &
Pólya inverse-gamma augmentation variable \\

$\gamma_{\mathrm E}$ &
Euler--Mascheroni constant \\

$\boldsymbol{\Psi}$ &
Collection of model parameters \\

\addlinespace[6pt]
\multicolumn{2}{l}{Variational quantities}\\
\addlinespace[3pt]

$q(\cdot)$ & Variational distribution \\

$\boldsymbol{\xi}$ &
Collection of variational parameters \\

$\mathbf a_i$ &
Variational mean vector of $\mathbf u_i$ \\

$\mathbf A_i$ &
Variational covariance matrix of $\mathbf u_i$ \\

$S_i$ &
Rate parameter of the Gamma variational
distribution for $\phi_i$ \\

$\widetilde{\eta}_{ij}$ &
Linear predictor evaluated at the variational
mean $\mathbf a_i$ \\

$\widetilde{\theta}_{ij}$ &
Variational expectation of $\theta_{ij}$ \\

$\widetilde c_{ij}$ &
Square root of the variational second moment
of $\theta_{ij}$ \\

$\widetilde{\pi}_{ij}$ &
Variational probability of membership in the
structural-zero component \\

$\widetilde z_{ij}$ &
Binary structural-zero assignment used in
classification VA \\

$\tau$ &
Classification threshold \\

$\mathcal L_{\mathrm{ZIDM}}$ &
Variational objective for the ZIDM-LVM \\

$\mathcal L_{\mathrm{ZIMN}}$ &
Variational objective for the ZIMN-LVM \\

\addlinespace[6pt]
\multicolumn{2}{l}{Simulation and evaluation quantities}\\
\addlinespace[3pt]

$R$ & Number of simulation replicates \\

$r$ & Simulation replicate index \\

$\vartheta$ &
Generic true parameter used in simulation
performance evaluation \\

$\hat{\psi}^{(r)}$ &
Estimate of $\vartheta$ in simulation replicate $r$ \\

$\hat{\rho}_{ij}$ &
Estimated underlying compositional proportion \\

$D_{\mathrm{Proc}}$ &
Procrustes error for ordination recovery \\

\end{longtable}

\normalsize

\subsection{Abbreviations}

\begin{table}[htbp]
\centering
\caption{Abbreviations used throughout the paper.}
\label{tab:abbreviations}
\small
\renewcommand{\arraystretch}{1.15}

\begin{tabular}{ll}
\toprule
\textbf{Abbreviation} & \textbf{Description} \\
\midrule

CI & Confidence interval \\

CLR & Centered log-ratio \\

CoDA & Compositional data analysis \\

DM & Dirichlet--multinomial \\

ELBO & Evidence lower bound \\

GLLVM & Generalized linear latent variable model \\

HTS & High-throughput sequencing \\

KL & Kullback--Leibler divergence \\

LVM & Latent variable model \\

MCMC & Markov chain Monte Carlo \\

MN & Multinomial \\

P-IG & Pólya inverse-gamma \\

RMSE & Root mean squared error \\

TMB & Template Model Builder \\

VA & Variational approximation \\

ZI & Zero inflation \\

ZIDM-LVM & Zero-inflated Dirichlet--multinomial latent variable model \\

ZIDM-K & Bayesian zero-inflated Dirichlet--multinomial method \\

ZIMN-LVM & Zero-inflated multinomial latent variable model \\

ZINB & Zero-inflated negative binomial \\

ZIPPCA & Zero-inflated probabilistic principal component analysis \\

ZR & Zero replacement \\

\bottomrule
\end{tabular}
\end{table}

\FloatBarrier

\bibliographystyle{apalike}
\bibliography{references}

\end{document}